%% file: crypto-verification.tex
\newif\ifdraft\draftfalse
\newif\iffull\fullfalse
\newif\ifsubmission\submissionfalse
\makeatletter \@input{texdirectives} \makeatother

\documentclass[
  \ifdraft draft,\else\fi
  acmsmall,10pt,
  \ifsubmission review,anonymous,\fi
  natbib=false]{acmart}
\acmConference[PL'18]{ACM SIGPLAN Conference on Programming Languages}{January 01--03, 2018}{New York, NY, USA}
\acmYear{2018}
\acmISBN{} 
\acmDOI{} 
\startPage{1}

\setcopyright{none}

\usepackage{booktabs}   
\usepackage{subcaption} 
\usepackage{fancyvrb}

\usepackage{amsmath,amsthm,mathpartir}
\usepackage{newtxmath}
\expandafter\let\csname not=\endcsname\relax
\expandafter\let\csname not<\endcsname\relax
\expandafter\let\csname not>\endcsname\relax
\usepackage{hyperref}
\ifdraft
\hypersetup{draft}
\fi
\usepackage{cleveref}
\usepackage{tikz}
\pgfdeclarelayer{background}
\pgfdeclarelayer{foreground}
\pgfsetlayers{background,main,foreground}
\usepackage[
  datamodel=acmdatamodel,
  style=acmauthoryear,
  backend=biber
]{biblatex}
\usepackage{xspace}
\usepackage{iris}

\newcommand{\nsSess}{\mathtt{sess}}

\usetikzlibrary{positioning,matrix,fit}

\newcommand{\vf}[1]{\mathit{#1}} \newcommand{\cf}[1]{\mathsf{#1}}
\newcommand{\tf}[1]{\mathtt{#1}}

\newcommand*{\KEY}{\cf{key}} \newcommand*{\key}[2]{\KEY_{#1}~#2}

\newcommand*{\MKAENCKEY}{\cf{mk\_aenc\_key}}
\newcommand*{\mkaenckey}{\MKAENCKEY}

\newcommand*{\MKSIGNKEY}{\cf{mk\_sign\_key}}
\newcommand*{\mksignkey}{\MKSIGNKEY} \newcommand*{\SIGNKEY}{\cf{sign\_key}}
\newcommand*{\signKey}[2]{\SIGNKEY~#1~#2}

\newcommand*{\INITSHARE}{\cf{init\_share}}

\newcommand*{\RESPSHARE}{\cf{resp\_share}}

\DeclareMathOperator{\mkNonce}{\cf{mk\_nonce}}

\newcommand*{\meta}[3]{#1,#2\mapsto#3}
\newcommand*{\TOKEN}{\cf{token}}
\newcommand*{\token}[2]{\TOKEN\,#1\,#2}

\newcommand*{\CONNPRED}{\cf{conn\_pred}}
\newcommand*{\connPred}[2]{\CONNPRED\;{#1}\;{#2}}
\newcommand*{\RPCPRED}{\cf{rpc\_pred}}
\newcommand*{\rpcPred}[2]{\RPCPRED\;{#1}\;{#2}} 
 
\DeclareMathOperator{\dhExp}{\char`\^}

\newcommand*{\PUBLIC}{\cf{public}} \newcommand*{\PUBLICKEY}{\cf{public\_key}}
\newcommand*{\COMPROMISED}{\cf{compromised}}

\newcommand*{\SECRET}{\cf{secret}}

 \newcommand*{\public}[1]{\PUBLIC\;{#1}}
\NewDocumentCommand\secret{O{} m}%
{\SECRET_{#1}\spac{#2}}

 \newcommand*{\SIGN}{\cf{sign}}
\newcommand*{\sign}[2]{\SIGN\,#1\,#2} \newcommand*{\VERIFY}{\cf{verify}}
\newcommand*{\verify}[2]{\VERIFY\,#1\,#2} 
\newcommand*{\OPEN}{\cf{open}} \newcommand*{\AENC}{\cf{aenc}}
\newcommand*{\ADEC}{\cf{adec}} \newcommand*{\PKEY}{\cf{pkey}}
\newcommand*{\SENC}{\cf{senc}} \newcommand*{\SDEC}{\cf{sdec}}

 \newcommand*{\open}[2]{\OPEN~#1~#2}
\newcommand*{\aenc}[2]{\AENC~#1~#2} \newcommand*{\adec}[2]{\ADEC~#1~#2}
\newcommand*{\senc}[2]{\SENC~#1~#2} \newcommand*{\sdec}[2]{\SDEC~#1~#2}
\newcommand*{\pkey}[1]{\PKEY~#1}

 \DeclareMathOperator{\Some}{\cf{Some}}
 \DeclareMathOperator{\None}{\cf{None}}

 \DeclareMathOperator{\send}{\cf{send}}
\DeclareMathOperator{\recv}{\cf{recv}}

 \DeclareMathOperator{\init}{\cf{init}}
\DeclareMathOperator{\resp}{\cf{resp}}

 \newcommand*{\pk}{\mathit{pk}}
\newcommand*{\sk}{\mathit{sk}}

\newcommand*{\session}{\cf{session}}

 \newcommand*{\termOwn}[3]{\boxedassert[densely
  dashed]{#3}[#1]_{#2}}

\newcommand{\cDbDisconnected}{\cf{DB.disconnected}}
\newcommand{\cDbConnected}{\cf{DB.connected}}

\newcommand{\cDbConnect}{\cf{DB.connect}}
\newcommand{\cDbClose}{\cf{DB.close}}
\newcommand{\cDbLoad}{\cf{DB.load}}
\newcommand{\cDbStore}{\cf{DB.store}}
\newcommand{\cDbCreate}{\cf{DB.create}}

\newcommand{\cConnect}{\cf{connect}}

\newcommand{\cLoad}{\cf{load}}
\newcommand{\cStore}{\cf{store}}
\newcommand{\cCreate}{\cf{create}}

\newcommand{\cClose}{\cf{close}}

\newcommand{\cdb}{\cf{db}}
\newcommand{\vdb}{\vf{db}}
\newcommand{\cConnConnected}{\cf{Conn.connected}}

\def\dbState{\cf{db\_state}}
\def\dbMain{\cf{db\_main}}
\def\dbCopy{\cf{db\_copy}}
\def\dbUpdate{\cf{db\_update}}
\def\dbSync{\cf{db\_sync}}
\def\client{\cf{client}}
\def\server{\cf{server}}
\def\isMap{\cf{is\_map}}
\newcommand*{\DbMainUpdate}{{\sc DbMainUpdate}}
\newcommand*{\DbCopyUpdate}{{\sc DbCopyUpdate}}
\newcommand*{\DbMainSync}{{\sc DbMainSync}}
\newcommand*{\DbStateAgree}{{\sc DbStateAgree}}
\newcommand*{\DbStateUpdate}{{\sc DbStateUpdate}}
\newcommand*{\DbMainAlloc}{{\sc DbMainAlloc}}
\newcommand*{\DbCopyAlloc}{{\sc DbCopyAlloc}}

\usepackage[margin=false,inline=true]{fixme}
\FXRegisterAuthor{aaa}{anaaa}{\color{cyan}AAA}
\FXRegisterAuthor{mg}{anmg}{\color{red}MG}
\FXRegisterAuthor{aja}{anaja}{\color{orange}AJA}
\newcommand{\aaa}[1]{\aaanote{#1}}

\theoremstyle{plain}
\newtheorem{theorem}{Theorem}[section]

\theoremstyle{definition}

\theoremstyle{remark}

\usepackage[final]{listings}
\begin{document}

\title{Cryptis: Cryptographic Reasoning in Separation Logic}


\author{Arthur Azevedo de Amorim} %
\affiliation{%
  \institution{Rochester Institute of Technology}
  \country{USA}
}

\author{Amal Ahmed}
\affiliation{%
  \institution{Northeastern University}
  \country{USA}
}

\author{Marco Gaboardi}
\affiliation{%
  \institution{Boston University}
  \country{USA}
}

\begin{abstract}
  We introduce \emph{Cryptis}, an extension of the Iris separation logic for the
  symbolic model of cryptography.  The combination of separation logic and
  cryptographic reasoning allows us to prove the correctness of a protocol and
  later reuse this result to verify larger systems that rely on the protocol.
  To make this integration possible, we propose novel specifications for
  authentication protocols that allow agents in a network to agree on the use of
  system resources.  We evaluate our approach by verifying various
  authentication protocols and a key-value store server that uses these
  authentication protocols to connect to clients.  Our results are formalized in
  Rocq.
\end{abstract}



\maketitle

\section{Introduction}
\label{sec:introduction}

Computer systems rely on various resources, such as IO devices, shared memory,
cryptographic keys or network connections.  The proper management of these
resources is essential to ensure that each system behaves correctly, in
particular in regards to security and privacy.  However, enforcing this
discipline is nontrivial, especially when resources are shared by multiple
components that might interfere with each other.  For example, a networked
system might rely on cryptographic protocols to secure its connections, and if
private keys are not shared between different components with care, the security
of the overall system may get compromised.
Good tool support can rule out potential conflicts in the use of shared
resources, making the system more reliable and secure.

A great tool for reasoning about resources is \emph{separation
  logic}~\cite{Reynolds02,Brookes07,OHearn07,BrookesO16}.  Assertions in
separation logic denote the ownership of resources, and if a program meets a
specification, it is guaranteed not to affect any resources that are disjoint
from those mentioned in its pre- or postconditions.  The notion of disjointness
is embodied by a special connective, the separating conjunction, that asserts
that multiple resources can be used independently, without conflict.  What
constitutes a resource and a conflict depends on the application at hand.
Originally~\cite{Reynolds02}, the resources were data structures in memory, and
the separating conjunction guaranteed the absence of aliasing.  In modern
versions of the logic, this has been generalized to cover other types of
resources, such as the state of a concurrent protocol~\cite{HinrichsenBK20} or
sources of randomness~\cite{BartheHL20,BaoGHT22,Li0H23}.

By describing precisely what resources each component can use, and how they are
used, separation logic brought a key advancement to program verification:
\emph{compositionality}. We can verify each component in isolation, without
knowing exactly what other resources might be used elsewhere. Later, we can
argue that the entire system is correct, provided that the resources used by
each component are separate at the beginning of the execution.  This allows the
logic to scale to large systems, including many that were challenging to handle
with prior techniques, such as concurrent or distributed ones.  And thanks to
its rich specification language, the logic can be used to reason about a wide
range of components with diverse purposes.  Individual proofs of correctness can
be composed in a unified formalism, thus ruling out bugs due to possible
mismatches between the guarantees of one component and the requirements of
another.

Due to the relative novelty of separation logic, however, the power of such
compositional reasoning remains underexplored in many domains.  Among many
examples, we can mention \emph{cryptographic protocols}.
To illustrate the issue, suppose that we would like to verify a distributed
application, such as a multi-threaded server that handles client requests
concurrently.
Several frameworks have been introduced over the years for tackling this task
under increasingly realistic assumptions~\cite{Krogh-Jespersen20, SergeyWT18,
  HinrichsenBK20}.
For example, \textcite{GondelmanHPTB23} showed how to verify an RPC library and
a key-value store on top of the Aneris logic~\cite{Krogh-Jespersen20}, which
assumes that messages can be dropped or duplicated, but not tampered with.
It would be desirable to extend these results to an even weaker model, where the
network is controlled by powerful attackers that can also forge messages or
tamper with them, and where reliable communication can only be enforced through
cryptographic protocols.
However, while several techniques have been developed to reason about such
protocols~\cite{Blanchet01, Bhargavan21, Vanspauwen015, ArquintSM023,
  DattaM0S11, MeierSCB13}, they have never been applied to reason about
application-level guarantees, such as proving that a client of our hypothetical
server only receives correct responses to its requests.

The goal of this paper is to show how we can connect these two lines of work.
We introduce a new logic called \emph{Cryptis}, which extends the Iris
separation logic~\cite{JungKJBBD18} to encompass the symbolic model of
cryptography.
Cryptis allows us to reason about the correctness of networked applications even
in the presence of powerful Dolev-Yao adversaries, which control the network but
cannot break cryptography.
The logic combines several ideas from the protocol-verification literature with
new features that allow verified protocols to be reused within larger systems,
where their integrity and confidentiality guarantees are crucial to ensure
correctness.

\paragraph*{Core features}

To reason about protocols in the presence of a Dolev-Yao attacker, Cryptis
follows prior work and uses a special $\PUBLIC$ predicate to overapproximate the
set of messages that the attacker can
access~\cite{Bhargavan21,Vanspauwen015,ArquintSM023}.  When a message is built
using cryptographic primitives, such as encryption or digital signatures,
Cryptis allows defining which properties must hold of the contents of the
message on a per-protocol basis.  These properties, which are guaranteed to hold
when messages are received, are instrumental to prove that protocols meet their
desired specifications.

To enable proof reuse when reasoning about larger systems, Cryptis associates
\emph{tokens} with cryptographic terms such as nonces, ephemeral session keys or
long-term private keys.  Tokens are separation-logic resources that can be used
to bind a term to metadata or other resources.  For example, when clients and
servers authenticate with each other, they can use metadata associated with the
exchanged session key to keep track of how many messages have been sent and
received through the connection, which allows them to transfer resources via
messages using the escrow pattern~\cite{KaiserDDLV17}, similar to what is done
in Aneris~\cite{Krogh-Jespersen20,GondelmanHPTB23}.  Because the connections are
authenticated, a server can assume that the exchanged resources pertain to a
specific client, which provides the capability to modify data belonging to that
client without interfering with others.

\paragraph*{Evaluation}

We evaluate Cryptis by verifying a series of components: a key-value store
server and client, an RPC mechanism, a reliable-connection abstraction, and
authenticated key-exchange protocols.  Our evaluation is carried out in a
modular fashion, where each component is built by reusing more basic
functionalities and verified solely based on their specifications, without
needing to access their implementation.  The proof of correctness of our
key-value store guarantees that clients always receive the expected values back
from the servers, implying that the integrity of their data is preserved.  To
the best of our knowledge, this is the first correctness proof for such a
distributed application that assumes a Dolev-Yao network.

Including cryptographic protocols in the model of a system also allows us to
analyze how the behavior of the system is affected when attackers can compromise
some honest agents---a common concern in modern
protocols~\cite{Cohn-GordonCG16}.  For example, we can prove that our key-value
store behaves correctly even when the agents' long-term keys are leaked to the
attacker, provided that the client communicates with the server using a session
key that was exchanged before the leak.

\paragraph*{Game-based specifications}

It is common to define the security of a cryptographic protocol via a
\emph{game}---a piece of code where honest agents aim to achieve some goal, such
as exchanging an unguessable session key, even in the presence of an attacker.
The protocol is secure, by definition, if the attacker cannot win the
game---that is, prevent the agents from achieving their goal.

Games are one of the main paradigms of specification in the computational model
of cryptography, where messages are bit vectors and adversaries are
probabilistic algorithms.  They provide an intuitive way to formulate concepts
such as ``secrecy'' that would be otherwise difficult to define.  Despite this
appeal, reasoning about games in the computational model is notoriously
difficult, because it requires intricate probabilistic arguments, needs to
account for multiple executions simultaneously, and usually involves reductions
(``if I can find an adversary against this protocol $P$, I can build an
adversary to break some problem $P'$ that is believed to be hard'').

By contrast, protocol specifications in Cryptis are formulated using a
specialized assertion language, which has built-in notions such as public terms
and metadata tokens.  Such specifications are much easier to prove than
analogues in the computational model, but it might not be clear what protection
they provide.  To clarify this point, we advocate for a methodology based on
\emph{symbolic} security games, a novel notion we introduce.  Like games for the
computational model, symbolic games are simply a piece of code where honest
agents interact with an attacker.  Their proofs of security, however, are much
simpler than reasoning about their counterparts in the computational model,
since they can be carried within the Cryptis logic.  The adequacy of the logic
allows us to translate such proofs to self-contained trace properties about the
operational semantics of games, which can be assessed independently of the
Cryptis assertion language.

\paragraph*{Trace-based specifications}

Verification tools for the symbolic model often define the correctness of a
protocol in terms of a global trace of events---ghost data that describes the
belief or the intent of each agent at various moments~\cite{Bhargavan21,
  Lowe97a, Blanchet02, ArquintSM023, MeierSCB13}.  For example, when an agent
completes an authentication handshake, they might produce an event to record
which nonces were exchanged and who they believe the other participant is.  We
can rule out various bugs in protocols by asking that these events correspond
somehow. To prevent a man-in-the-middle attack, we can verify that the event
marking the completion of a handshake is matched by an earlier event signaling
that an agent accepted to partake in the handshake; to prevent replay attacks,
we can strengthen this requirement so that the acceptance event occur at most
once for a given combination of nonces.

Cryptis follows a different approach.  Rather than relying on a global trace of
events that is hardwired into the formalism, users are free to plug in their own
ghost state when specifying protocols and reasoning about them---typically, by
combining ghost state with term metadata.  An authentication protocol, as we
will see, is simply a means for the agents to agree on a secret shared key and
establish ghost resources to coordinate their actions. We could use ghost state
to store an event trace, in which case it would be possible to adapt the
classical notions of authentication into Cryptis.  Nevertheless, we have not
found a reason to do so. As we will see, our specifications are capable of
preventing similar bugs compared to those based on event traces, with the
advantage that they do not require the same kind of temporal reasoning.

\paragraph*{Secrecy as a resource}

The combination of separation logic and cryptographic reasoning provides a fresh
look on how to reason about protocol security.  Cryptis allows us to model the
secrecy of a private key or other sensitive cryptographic terms with a
separation-logic resource $\secret{k}$.  While the resource is available, $k$ is
guaranteed to be secret, but we can consume the resource at any moment to
compromise $k$ and make it available to the attacker.

This enables a new model of dynamic compromise.  In systems based on a global
event trace~\cite{Bhargavan21,ArquintSM023}, the attacker API includes functions
that allow any agent or session to be compromised at any point.  Operationally,
this has the effect of adding a special ``compromise'' event to the global
trace, indicating the moment when the compromise occurred and allowing the
attacker to access private keys and other compromised data.  But because a
compromise can occur at any point, it is difficult to reason about the behavior
of a protocol under a specific compromise scenario (e.g., where a key is
compromised only after a certain event takes place).
Cryptis, on the other hand, allows us to model specific compromise scenario in
symbolic security games, by adding a command to leak sensitive cryptographic
material $t$ such as a private key in a specific step of the game.  If we keep
the secrecy resource associated with $t$ before the compromise happens, we can
argue that any actions that take place before the compromise are unaffected by
it.

\paragraph*{Summary of contributions}

\begin{itemize}
\item A new separation logic, \emph{Cryptis}, that extends
  Iris~\cite{JungKJBBD18} with the symbolic (or Dolev-Yao) model of cryptography
  (\Cref{sec:cryptis}).  The logic demonstrates how we can reason about rich
  protocol properties in the presence of powerful attackers and dynamic
  compromise without relying on a global trace of events.
\item Novel specifications for authentication protocols that allow agents to
  coordinate their actions via resources tied to cryptographic material
  (\Cref{sec:nsl,sec:iso}).
\item A new model of key compromise that treats the secrecy of a key as a
  separation-logic resource.  If a specification guarantees a certain property
  unless a key has been compromised, such a secrecy resource could be used to
  argue that the desired protocol outcome is met.
\item A novel methodology for assessing protocol security via \emph{symbolic
    security games}.
\item A demonstration that our specifications are expressive enough to reason
  about complex systems built on top of cryptographic protocols. We define a
  key-value store service that uses authenticated connections to communicate
  with clients and prove that the authenticated connections are enough to
  preserve the integrity of the clients' data.
\end{itemize}

Our main results are formalized in Rocq using the Iris separation
logic~\cite{JungKJBBD18}.

\paragraph*{Structure of the paper}

\Cref{sec:cryptis} gives a comprehensive view of the Cryptis logic.
In \Cref{sec:key-value-overview}, we present the architecture of a modular
key-value store cloud application and its specification in Cryptis.
In the remaining sections, we discuss how to implement and verify each component
of this store.
First, we discuss how to verify \emph{authentication protocols}, which allow
agents to exchange such symmetric encryption keys for establishing sessions.  We
present correctness proofs for the classic Needham-Schroeder-Lowe
protocol~\cite{NeedhamS78,Lowe96} (\Cref{sec:nsl}), which uses asymmetric
encryption, and the ISO protocol~\cite{Krawczyk03} (\Cref{sec:iso}), which uses
Diffie-Hellman key exchange and digital signatures.  For the latter, we show
that the protocol guarantees \emph{forward secrecy}: session keys remain secret
even after long-term keys are compromised.
The authentication protocols can be reused to prove the correctness of an
\emph{authenticated, reliable channel abstraction}
(\Cref{sec:reliable-connections}) which, in turn, can be used to implement an
remote procedure call mechanism (\Cref{sec:rpc}).  This mechanism allows the
key-value store server to communicate with its clients in a secure manner, thus
allowing us to prove its correctness (\Cref{sec:key-value}).
We implemented Cryptis in Rocq using Iris~\cite{JungKJBBD18}, an extensible
higher-order concurrent separation logic, and mechanized all our case studies
using the Iris proof mode~\cite{Krebbers0BJDB17} (\Cref{sec:implementation}).
The implementation and the case studies are included in the supplementary
material.
We discuss related work in \Cref{sec:related-work} and conclude in
\Cref{sec:conclusion}.

\section{Core Cryptis}
\label{sec:cryptis}

Cryptis is a logic for reasoning about networked programs in a mostly standard
functional imperative language.  The logic and its programming language are
summarized in \Cref{fig:cryptis}.  Most features are inherited from Iris, so we
will focus on the Cryptis extensions, and refer readers to
\textcite{JungKJBBD18} for more background on the other features.  The $\later$
symbol refers to the later modality of Iris, which is used to state recursive
definitions while avoiding paradoxes.  The assertion $\always P$ means that $P$
holds persistently, without holding any resources.  The assertion
$P \vs[\mask] Q$ means that we can make $Q$ hold by consuming the resource $P$,
modifying ghost state and accessing invariants under any namespace
$\namesp \in \mask$.

\begin{figure}
  \small
  \begin{align*}
    & \text{Key types} & u & := \AENC \mid \ADEC \mid \SIGN \mid \VERIFY \mid
                             \SENC \mid \cdots \\
    & \text{Functionalities} & F & := \AENC \mid \SIGN \mid \SENC \mid \cdots \\
    & \text{Terms} & t, \sk, \pk, k & := n \mid \namesp \mid (t_1, t_2) \mid \{t\}@k \mid
                            \key{u}{t} \mid
                            t \dhExp { (t_1 \cdots t_n) } \mid \cdots \\
    & \text{Expressions} & e & := \send{e} \mid \recv~() \mid
                               \{e\}@e \mid \key{u}{e} \mid \open{e_1}{e_2} \mid
                               \pkey{e} \mid e_1 \dhExp e_2 \\
                               & & & \mid \mkNonce~() \mid \cdots \\
    & \text{Assertions} & P, Q
    & := \underbrace{
      \meta{F}{\namesp}{\varphi} \mid
      \meta{t}{\namesp}{x} \mid
      \public{t}}_{\text{persistent}}
      \mid \token{F}{\mask} \mid \token{t}{\mask} \mid \cdots
  \end{align*}

  \begin{minipage}[t]{0.5\linewidth}%
    \centering
    Term equations
    \begin{align*}
      t \dhExp () & = t \\
      t \dhExp (t_1 \cdots t_n) \dhExp t_{n+1}
                  & = t \dhExp (t_1 \cdots t_{n+1}) \\
      t \dhExp (t_1 \cdots t_k t_{k+1} \cdots t_n)
                  & = t \dhExp (t_1 \cdots t_{k+1} t_k \cdots t_n)
    \end{align*}
  \end{minipage}%
  \begin{minipage}[t]{0.5\linewidth}%
    \centering
    Derived notions
    \begin{align*}
      \secret{t}
      & \eqdef
        \setlength{\arraycolsep}{0pt}
        \begin{array}[t]{l}
          (\public{t} \wand \later \FALSE) \land (\TRUE \vs \public{t}) \\
          {} \land (\TRUE \vs (\public{t} \wand \later \FALSE))
        \end{array} \\
      \termOwn{t}{\namesp}{a}
      & \eqdef \exists \gname, \meta{t}{\namesp}{\gname} * \ownGhost{\gname}{a}
    \end{align*}
  \end{minipage}

  \vspace{2em}

  \begin{minipage}[t]{0.7\linewidth}%
    \centering
    Public terms
    \begin{align*}
      \public{n}, \public{\namesp} & \iff \TRUE \\
      \public{(t_1,t_2)} & \iff \public{t_1} \land \public{t_2} \\
      \public{\{(\namesp, t_1)\}}@(\key{u}{t_2})
      & \iff
        \setlength{\arraycolsep}{0pt}
        \begin{array}[t]{l}%
          \public{t_1} \land \public{(\key{u}{t_2})} \\
          {} \lor \exists \varphi, \meta{F_u}{\namesp}{\varphi} * \always
          \varphi~(\key{u}{t_2})~t_1 \\
          \quad {} *
          \always (\public{(\key{\bar{u}}{t_2})} \wand \public{t_1})
        \end{array} \\
      \public{(\key{u}{t})}
      & \iff \public{t} \lor \PUBLICKEY~u \\
      \public{(t \dhExp t')}
      & \iff \TRUE \qquad (\text{when $t$ is not a DH term}) \\
      \public{(t \dhExp (t_1 \cdots t_n))}
      & \iff
        \setlength{\arraycolsep}{0pt}
        \begin{array}[t]{l}%
          \exists i, \public{(t \dhExp (t_1 \cdots t_{i-1} t_{i+1} \cdots
          t_n))} \\
          {} \land \public{t_i}
        \end{array} \\
    \end{align*}
  \end{minipage}%
  \hfill%
  \begin{minipage}[t]{0.3\linewidth}%
    \centering
    Key types
    \begin{align*}
      \overline{\AENC} & = \ADEC \\
      \overline{\SIGN} & = \VERIFY \\
      \overline{\SENC} & = \SENC.
    \end{align*}

    \vspace{1em}

    \begin{mathpar}
      \inferrule
      { u \in \{\AENC, \VERIFY\} }
      { \PUBLICKEY~u }
    \end{mathpar}
  \end{minipage}

  \vspace{2em}

  Operational semantics
  \begin{align*}
    \open{(\{t_1\}@(\key{u}{t_2}))}{(\key{\bar{u}}{t_2})}
    & \to \Some~t_1 \\
    \open{t}{k} & \to \None & &  \text{(in all other cases)}
  \end{align*}

  \vspace{2em}

  Program logic
  \begin{mathpar}
    \hoare{\later \public{t}}{\send{t}}{\TRUE} \qquad
    \hoare{\TRUE}{\recv~()}{t, \public{t}} \\
    \hoare{\forall t\,t', t' \in T(t) \wand t' \preceq t}{\mkNonce~()}{t,
    \always (\public{t} \iff \later \always \varphi~t) * \Sep_{t' \in T(t)}\token{t'}{\top}}
  \end{mathpar}

  \vspace{2em}

  Metadata rules (for both terms and tag invariants)
  \begin{minipage}[t]{0.4\linewidth}%
  \begin{align*}
    \meta{\alpha}{\namesp}{\beta_1} * \meta{\alpha}{\namesp}{\beta_2}
    & \vdash \later (\beta_1 = \beta_2) \\
    \upclose\namesp \subseteq \mask *
    \token{\alpha}{\mask}
    & \vs \meta{\alpha}{\namesp}{\beta}
  \end{align*}
  \end{minipage}%
  \begin{minipage}[t]{0.6\linewidth}%
  \begin{align*}
    \meta{\alpha}{\namesp}{\beta} * \token{\alpha}{\{\namesp\}}
    & \vdash \FALSE \\
    \token{\alpha}{(\mask_1 \uplus \mask_2)}
    & \vdash \token{\alpha}{\mask_1} * \token{\alpha}{\mask_2} \\
  \end{align*}
  \end{minipage}
  \caption{The Cryptis logic and programming language.}
  \label{fig:cryptis}
\end{figure}

There are two primitive networking functions, $\send$ and $\recv$.  These
functions are restricted to \emph{terms $t$}, a subset of values that excludes
anything that cannot be meaningfully sent over the network, such as pointers or
closures.  We use $k$ to range over terms that serve as keys, and reserve $\sk$
for private keys and $\pk$ for public keys. The metavariable $\namesp$ ranges
over namespaces, which we will use as tags to to distinguish different types of
messages that can arise on the network.

\paragraph{Cryptographic operations}

Terms can be manipulated with several cryptographic primitives: \emph{sealing}
($\{t\}@k$), \emph{Diffie-Hellman exponentiation} ($\dhExp$) and \emph{nonce
  generation} ($\mkNonce$).  Sealing is an umbrella primitive used to encode
various encryption-like functionalities. In $\{t\}@k$, the term $t$ is the
sealed message, and $k$ is the key used to seal it.  Keys are terms of the form
$\key{u}{t}$, where $t$ is the seeding material used to generate it and $u$ is
the type of the key.  We distinguish between keys for asymmetric encryption
($u \in \{\AENC, \ADEC\}$), digital signatures ($u \in \{\SIGN, \VERIFY\}$) and
symmetric encryption ($u = \SENC$).  We can unseal a sealed term by calling
$\OPEN$, which succeeds only if the key used for sealing matches the one used
for unsealing. The partial operation $\bar{u}$ maps a key type $u$ to its
corresponding opening key.  The expression $\pkey{\sk}$ computes the public key
corresponding to some secret key $\sk$.
In a Diffie-Hellman term $t \dhExp (t_1 \cdots t_n)$, the terms
$t_1, \ldots, t_n$ represent the exponents.  We quotient terms to validate all
the equations we want to hold of Diffie-Hellman terms; in particular, exponents
can be freely permuted, and we have the familiar identity
$t \dhExp t_1 \dhExp t_2 = t \dhExp t_2 \dhExp t_1$, which allows agents to
compute a shared Diffie-Hellman secret based on their key shares $t \dhExp t_1$
and $t \dhExp t_2$.
A call to $\mkNonce$ produces a fresh nonce that has never appeared in the
program state before.

For clarity, we will use different names to refer to sealing and opening in our
examples, depending on the cryptographic primitive we are modeling.  We use
$\aenc{\pk}{t}$ and $\adec{\sk}{t}$ in the case of asymmetric encryption,
$\sign{\sk}{t}$ and $\verify{\pk}{t}$ in the case of digital signatures, and
$\senc{k}{t}$ and $\sdec{k}{t}$ in the case of symmetric encryption.  Note that
$\VERIFY$ outputs the contents of the signed message, instead of a success bit.
We will also use the abbreviations
$\mkaenckey \eqdef \key{\ADEC}{(\mkNonce~())}$ and
$\mksignkey \eqdef \key{\SIGN}{(\mkNonce~())}$, which are used to generate new
secret keys for asymmetric encryption and digital signatures.

\paragraph{Attacker model}

Cryptis assumes a Dolev-Yao, or symbolic, model of cryptography, where
cryptographic operations behave as perfect black boxes. It is impossible to
manipulate messages directly as bit strings, to guess nonces or keys out of thin
air, by sampling, or to extract the contents of an encrypted message without the
corresponding key.  On the other hand, we assume that the network is controlled
by an attacker that has the power to drop, duplicate, or manipulate network
messages arbitrarily by applying cryptographic operations.  (We model this in
our operational semantics by having the $\send$ and $\recv$ functions
communicate with a special attacker thread that sits in a loop performing these
actions nondeterministically.)  As usually done in works based on the symbolic
model, messages do not include their sender or recipient, since this information
is not reliable.

\paragraph{Public terms}

To allow agents to communicate securely in the presence of such a powerful
attacker, Cryptis forces every message to travel through the network to satisfy
a special $\PUBLIC$ predicate.  Accordingly, the specification of the networking
functions says that $\send$ takes in a public term, whereas $\recv$ is
guaranteed to return a public term.  The definition of $\PUBLIC$ balances
between two needs: capturing the capabilities of the attacker, on the one hand,
and allowing honest agents to reason about their communication, on the other.

To capture the attacker capabilities, the definition ensures that public terms
are preserved by all term operations (pairing two terms, sealing a term with a
key, etc.).
To make it possible to reason about communication, the public predicate allows
us to impose invariants on sealed messages, as it is often done in similar
tools~\cite{Bhargavan21, BackesHM11, BackesHM14}.  Suppose that we want to send
a message of the form $\{(\namesp, t_1)\}@\pk$, where $\pk = \key{\AENC}{t_2}$
is an encryption key.  In typical uses of encryption, the contents of the
message, $t_1$, are usually not public.  Nevertheless, the definition of public
on sealed terms says that we can still prove that the message is public provided
that $t_1$ satisfies an invariant $\varphi$ attached to the tag $\namesp$.
The predicate $\meta{F_u}{\namesp}{\varphi}$ means that the tag $\namesp$ is
associated with the invariant $\varphi$ for sealed messages under the
functionality $F_u$.  Each tag can be associated with at most one invariant
under a given functionality.  The protocol verifier chooses which invariants to
use by consuming a $\token{F}{\mask}$ resource, which states that no tags in
$\mask$ have had any invariants assigned to them.  (Initially, $\mask$ is set to
$\top$, which contains every tag.)  Cryptis focuses on tagged messages to make
reasoning modular: if two protocols use different tags, their proofs can be
automatically composed.  The clause for encrypted terms also includes
$\always (\public{(\key{\bar{u}}{t_2})} \wand \public{t_1})$, which allows the
attacker to conclude that the contents of the message are public if they are
ever able to decrypt it with a public (compromised) decryption key.  If
$u = \SIGN$, then the antecedent of this implication is trivial, which means
that we can only sign messages when the contents are public.  Notice that both
this implication and the message invariant $\varphi$ are guarded by the
persistence modality $\always$: since a Dolev-Yao attacker can duplicate a
message arbitrarily, we want $\PUBLIC$ to be persistent.

Dually, when another agent receives a sealed message, the definition says that
we must consider two cases: either the sealing key and the contents of the
message are public, or the corresponding sealing invariant holds.  The first
case typically occurs when reasoning about communication with an attacker or
compromised agent, which cannot be expected to enforce non-trivial invariants,
whereas the second one arises when communicating with another honest agent.
This type of case analysis is common in the literature on protocol
verification~\cite{BackesHM11,BackesHM14}.

The definition of $\PUBLIC$ on keys allows them to be shared according to their
common usage patterns.  For asymmetric encryption, the sealing key ($u = \AENC$)
is always considered public, whereas the unsealing key ($u = \ADEC$) is public
if and only the seed is.  For digital signatures, it's the opposite.  A
symmetric encryption keys is public if and only if its seed is.

The definition of $\PUBLIC$ on Diffie-Hellman terms allows agents to freely
exchange key shares $t \dhExp t'$ in cleartext.  When more than one exponent is
involved, the term is public if and only if it can be obtained by combining two
smaller public terms via exponentiation.  In particular, if $t$ is not a DH
term, then $\public{t \dhExp t_1t_2} \iff \public{t_1} \lor \public{t_2}$.

Another example of terms with a non-trivial public predicate are nonces.  The
specification for the $\mkNonce$ function says, among other things, that the
resulting nonce $t$ is such that $\public{t}$ can be defined to be any predicate
$\varphi~t$ chosen at the moment of verification.  (The use of the later
modality $\later$ allows us to prevent contradictions when handling circular
definitions, such as $\varphi~t \eqdef \neg \varphi~t$.)  We can choose
$\varphi~t = \TRUE$ to generate a public nonce, or $\varphi~t = \FALSE$ to
generate a nonce that can never be disclosed to the attacker.

Another possibility is to defer the choice of this predicate.  The resource
$\secret{t}$ means that $t$ is not currently public, but can be made public or
permanently secret at any point. This resource can be obtained by setting
$\varphi~t$ to some persistent assertion $P$ such that
$\neg P \land (\TRUE \vs P) \land (\TRUE \vs \always \neg P)$ holds.  (This
assertion could be defined using a variety of ghost-state constructions,
including the term metadata feature described next.)  We can also create secrecy
resources for some terms derived from the nonce, such as a private key.  This
pattern is useful to model dynamic compromise scenarios, in which an attacker
does not have access to some key at first, but eventually manages to compromise
it.

\paragraph{Term metadata}

The last feature that we need to cover is \emph{term metadata}.  The assertion
$\meta{t}{\namesp}{x}$ says that the term $t$ has been permanently associated
with the metadata item $x$ under the namespace $\namesp$.  The metadata $x$
ranges over elements from arbitrary countable types.  Similarly to sealing
predicates, we can create such an association by consuming a suitable resource
$\token{t}{\mask}$.  Each term is associated to at most one metadata item under
a given namespace $\namesp$.  Metadata tokens are created during nonce
generation.  As shown in the specification for $\mkNonce$, the post-condition
contains tokens for any term $t' \in T(t)$, where $T(t)$ is a finite set that is
required to list terms $t'$ that mention $t$ as a subterm (written
$t \preceq t'$).

Metadata in Cryptis serves multiple purposes.  One purpose is to reason about
term freshness: if a set of terms $T$ is such that every term has defined
metadata under some namespace $\namesp$, any term $t$ that still has a token
containing $\namesp$ must not belong to $T$.
More interestingly, metadata allows different agents on a network to coordinate
their actions.  The derived construct $\termOwn{t}{\namesp}{a}$ allows us to
associate a term $t$ with ghost state $a$ drawn from arbitrary resource
algebras.  This will be useful when reasoning about the state of a connection or
other system components, as we will see.

\section{Motivating application: a key-value store}
\label{sec:key-value-overview}

\begin{figure}[t]
  \centering
  \input{figures/system-structure}
  \caption{Structure of key-value store.  Arrows denote dependencies, circles
    denote internal components, and squares denote the key-value store itself.}
  \label{fig:key-value-structure}
\end{figure}

In the rest of the paper, we will illustrate the expressiveness of Cryptis by
verifying the correctness of a simple key-value store.  This case study
demonstrates how we can reason modularly about a high-level application that
provides non-trivial integrity guarantees even in the presence of arbitrary
Dolev-Yao attackers.  It also showcases how these guarantees can degrade in the
presence of an attack.  In this section, we content ourselves with an overview
of the architecture of the key-value store and its specification.  Later, we
will dive into its individual components.

\Cref{fig:key-value-structure} summarizes the structure of the application. A
server stores client data in internal data structures, and the client performs
API calls to retrieve and manipulate its data.  The communication between the
client and the server is implemented by a remote procedure call (RPC) component,
which sits on top of a connection abstraction that preserves the ordering and
contents of messages.  To create a connection, the client must initiate an
authentication handshake with the server, which allows the two parties to
exchange a session key and confirm each other's identities.  Since our focus is
on how such distinct components can be developed and verified modularly, the
functionality of each verified component will be rather minimal.  For example,
the store server provides sequential consistency, runs on a single machine
machine, and stores the client data using an association list.  Because of the
modular design and the expressiveness of separation logic, it should be possible
to make each component more realistic without changing fundamentally how they
are connected.

\Cref{fig:key-value-spec} shows the specification of the client API.  To
interact with the server, the client must first connect to it using the
$\cConnect$ function.  This function returns a connection object $c$, together
with a resource $\cDbConnected~\sk_C~\sk_S~c$ that indicates that the client is
connected.  While the client is connected, it can perform operations on its
data: load a value stored under a key ($\cLoad$), store a value under a new key
($\cCreate$), or update the value of an existing key ($\cStore$).  The
specifications are modeled after the specifications for memory operations in
separation logic, with one minor twist: a load can return an incorrect value if
the connection is compromised.  A connection is compromised if either the server
or the client was compromised when the connection was established (that is, if
their private keys were known to the attacker).

\begin{figure}[t]
  \centering
  \small
  \begin{tabular}[t]{c|l}
    Notation & Meaning \\ \hline
    $\sk_C$ & Client's secret key \\
    $\sk_S, \pk_S$ & Server's secret key and public keys \\
    $k_c$ & Session key used in connection $c$ \\
    $\cDbDisconnected~\sk_C~\sk_S$ & The client is disconnected from the server \\
    $\cDbConnected~\sk_C~\sk_S~c$ & The client is connected to the server via
                                    the connection $c$ \\
    $\COMPROMISED~\sk_C~\sk_S~c$ & The connection $c$ is compromised and accessible to the
                       attacker \\
    $T \mapsto^{\sk_C,\sk_S}_{\cdb} \bot$ & No term $t \in T$ is stored in the server \\
    $t_1 \mapsto^{\sk_C,\sk_S}_{\cdb} t_2$ & The value $t_2$ is stored under the
                                             key $t_1$
  \end{tabular}
  \vspace{2em}
  \begin{mathpar}
    \hoare
    { \cDbDisconnected }
    { \cDbConnect~\sk_C~\pk_S }
    { c, \cDbConnected~c }

    \and

    \hoare
    { \cDbConnected~c }
    { \cDbClose~c }
    { \cDbDisconnected * \public{k_c} }

    \and

    \hoare
    { \cDbConnected~c * t_1 \mapsto_{\cdb} t_2 }
    { \cDbLoad~c~t_1 }
    { t_2', (\COMPROMISED~c \lor t_2' = t_2) * \cDbConnected~c
      * t_1 \mapsto_{\cdb} t_2 }

    \and

    \hoare
    { \cDbConnected~c * t_1 \mapsto_{\cdb} \bot }
    { \cDbCreate~c~t_1~t_2 }
    { \cDbConnected~c * t_1 \mapsto_{\cdb} t_2 }

    \and

    \hoare
    { \cDbConnected~c * t_1 \mapsto_{\cdb} t_2' }
    { \cDbStore~c~t_1~t_2 }
    { \cDbConnected~c * t_1 \mapsto_{\cdb} t_2 }

    \and t_1 \mapsto_{\cdb} t_2 * t_1 \mapsto_{\cdb} t_2' \vdash \FALSE

    \and \secret{\sk_C} * \secret{\sk_S} * \cDbConnected~c
    \vdash \later \always \neg \COMPROMISED~c

    \and
    \token{\sk_C}{(\upclose{\namesp_{\cdb}.\client.\sk_S})}
    \vs[\top]
    \cDbDisconnected * \top \mapsto_{\cdb} \bot

    \and

    T_1 \uplus T_2 \mapsto_{\cdb} \bot
    \vdash T_1 \mapsto_{\cdb} \bot * T_2 \mapsto_{\cdb} \bot

  \end{mathpar}
  \caption{Key-value store assertions and specifications for the client API.
    For readability, the specifications omit some of the key parameters that
    appear in the predicates. The terms $t_1$, $t_2$ and $t_2'$ are assumed
    public.}
  \Description{Key-value store specification}
  \label{fig:key-value-spec}
\end{figure}

We can rule out the possibility of a compromise by proving that the agents'
private keys were still secret at any point after the connection $c$ was
established.  Note that it is still possible that these private keys end up
leaking at a later point without affecting the integrity of $c$.  As we will
see, this is a product of the post-compromise guarantees of the underlying
key-exchange protocol. Logically, this is a consequence of the way persistent
assertions work in Iris.  The assertion
$\later\always\neg\COMPROMISED~\sk_C~\sk_S~c$ is persistent because it is
guarded by the persistence modality $\always$.  This means that, if we are in a
proof context where $\secret{\sk_C}$, $\secret{\sk_S}$ and
$\cDbConnected~\sk_C~\sk_S~c$ all hold, we can prove
$\later\always\neg\COMPROMISED~\sk_C~\sk_S~c$ without consuming these premises.
Later, we can consume $\secret{\sk_C}$ or $\secret{\sk_S}$ to leak one of the
private keys.

At any moment, the client can choose to disconnect from the server by calling
the $\cClose$ function.  After disconnecting, $k_c$, the session key used to
encrypt the connection, is no longer needed, so it can be made public and leaked
to the attacker.  Of course, if this were the case, the attacker would be able
to read any messages that were encrypted with it, ruining any confidentiality
guarantees.  However, such confidentiality guarantees are relational properties
that are currently out of reach for Cryptis, which is a unary logic---indeed, we
assume that the database contents are always public.  We allow the session key
to be leaked after disconnection to highlight that this would not affect the
\emph{integrity} of the client's data: the attacker could try to send requests
to the server using the compromised key, but those requests would be ignored.

To illustrate how these specifications might be used, let us assess the
integrity of our key-value store using a simple security game
(\Cref{fig:key-value-game}).  The game sets up digital signature keys for the
client and the server, sends the long-term public keys to the attacker, and then
runs the client and the server in parallel. The client uses the server to store
a key-value pair that is chosen by the attacker and then tries to retrieve that
value from the server.
Our goal is to prove that the client's assertion succeeds; that is, the client
reads back the same value that it stored originally.  Moreover, this assertion
succeeds even though various keys are leaked during the game.
Though the game code is simple, its operational semantics is subtly complex: the
agents run concurrently and their interaction is mediated by a Dolev-Yao
attacker.  Thus, checking the security of the game forces us to reason about
concurrency, making it challenging to provide similar formulations in systems
that do not cover this feature, such as DY*~\cite{Bhargavan21}.

To prove that the game is secure, we use the specification of the $\mkNonce$
function to generate the signature keys $\sk_C$ and $\sk_S$ so that we obtain
secrecy resources $\secret{\sk_C}$ and $\secret{\sk_S}$.  We also obtain
metadata tokens for these keys, which we can use to initialize the ghost state
required to run the server and the client.  For the client, this means obtaining
the assertions $\cDbDisconnected~\sk_C~\sk_S$ and
$\top \mapsto^{\sk_C,\sk_S}_{\cdb} \bot$, which guarantees that the database is
currently uninitialized.  (The lemmas used to initialize the server's ghost
state are omitted for brevity.)

We pass all these resources to the \verb!run_client! function.  When we close
the first connection $c_1$, we are allowed to leak its session key thanks to the
specification of $\cDbClose$.  When we establish the second connection $c_2$, we
use $\secret{\sk_C}$ and $\secret{\sk_S}$ to prove that $c_2$ is not
compromised, which allows us to prove that $\cDbLoad$ returns the expected
value.

\begin{figure}[t]
  \centering
\small
\begin{minipage}[t]{0.4\linewidth}%
\vspace{0pt}
\begin{BVerbatim}
let run_client leak_keys skC pkS =
  let c1 = DB.connect skC pkS in
  let key = recv () in
  let val = recv () in
  DB.create c1 key val;
  DB.close c1;
  send (session_key c1);

  let c2 = DB.connect skC pkS in
  leak_keys ();
  let val' = DB.load c2 key in
  assert (val = val')
\end{BVerbatim}
\end{minipage}\hfill%
\begin{minipage}[t]{0.55\linewidth}%
\vspace{0pt}
\begin{BVerbatim}
let game () =
  let skC, skS =
    mk_sign_key (), mk_sign_key () in
  let pkC, pkS = pkey skC, pkey skS in
  send pkC; send pkS;

  let leak_keys () = send skC; send skS in

  fork (fun () -> DB.start_server skS);
  fork (fun () -> run_client leak_keys skC pkS)
\end{BVerbatim}
\end{minipage}
\caption{Security game for the key-value storage service.  The client reads back
  the value they stored even if long-term keys are leaked after the connection.}
\Description{Security game for the key-value storage service.}
\label{fig:key-value-game}
\end{figure}

\section{Authentication: the NSL protocol}
\label{sec:nsl}

The first component of the key-value store we will analyze is its authentication
protocol.  To guarantee post-compromise security, the implementation of the
store uses a protocol based on Diffie-Hellman key exchange, which we will cover
in \Cref{sec:iso}.  Before we do so, however, we will consider the
Needham-Schroeder-Lowe protocol~\cite{NeedhamS78,Lowe96} (\emph{NSL}), a classic
protocol based on public-key encryption.  Though the protocol provides weaker
security guarantees, it is often used as an introductory example in
protocol-verification tools and, thus, serves as a good point of comparison for
Cryptis.
There are two versions of the protocol: one that relies on a trusted server to
distribute public keys, and one where the participants know each other's public
keys from the start.  For simplicity, we model the second one.  A typical run
can be summarized as follows:
\begin{align*}
  I \to R & : \aenc{\pk_R}{[\cf{m1}; a; \pk_I]} &
  R \to I & : \aenc{\pk_I}{[\cf{m2}; a; b; \pk_R]} &
  I \to R & : \aenc{\pk_R}{[\cf{m3}; b]}.
\end{align*}
First, the initiator $I$ generates a fresh nonce $a$ and sends it to the
responder $R$, encrypted under their public key $\vf{pk}_R$.  The responder
replies with $a$ together with a fresh nonce $b$.  The initiator confirms the
end of the handshake by returning $b$.  If the protocol terminates successfully,
and both agents are honest, they can conclude that their identities are
correct---that is, they match the public keys sent in the messages---and that
the nonces $a$ and $b$ are secret.  In particular, they can use $a$ and $b$ to
derive a secret session key to encrypt further communication.  \Cref{fig:nsl}
shows an implementation of the protocol in the Cryptis programming language.
To keep examples short, we'll use a syntax inspired by the ProVerif protocol
analyzer~\cite{Blanchet01}: \verb!let!  declarations can mention patterns of the
form \verb!=p!, which are only matched by \verb!p! itself.
Any errors that arise during execution, such as failed pattern matching, cause
the code to safely return \verb!None!.  (Formally, these errors are managed
using the option monad, and \verb!let! in our code snippets should be read as
monadic bind.)

\begin{figure}[t]
  \centering
\small
  \begin{minipage}[t]{0.5\linewidth}
\vspace{0pt}
\begin{BVerbatim}
let initiator skI pkR =
  let pkI = pkey skI in
  let a = mk_nonce () in
  let m1 = aenc pkR ["m1"; a; pkI] in
  send m1;
  let m2 = recv () in
  let [="m2"; =a; b; =pkR] = adec skI m2 in
  let m3 = aenc pkR ["m3"; b] in
  send m3;
  let k = key senc [pkI; pkR; a; b] in
  k
\end{BVerbatim}
\end{minipage}%
\begin{minipage}[t]{0.5\linewidth}%
\vspace{0pt}
\begin{BVerbatim}
let responder skR =
  let pkR = pkey skR in
  let m1 = recv () in
  let [="m1"; a; pkI] = adec skR m1 in
  if not (is_aenc_key pkI) then fail ();
  let b = mk_nonce () in
  let m2 = aenc pkR ["m2"; a; b; pkR] in
  send m2;
  let m3 = recv () in
  let [="m3"; =b] = adec skR m3 in
  let k = key senc [pkI; pkR; a; b] in
  (pkI, k)
\end{BVerbatim}
\end{minipage}
\caption{Implementation of the NSL protocol. Variables beginning with
  \texttt{sk} and \texttt{pk} refer to secret and public keys.}
  \Description{Implementation of the NSL protocol}
  \label{fig:nsl}
\end{figure}

\subsection{Proving security}

The NSL handshake produces a session key $k$ that is guaranteed to be secret, as
long as both participants are honest.  We formalize this claim with the
following theorem, which, moreover, produces metadata tokens for the agents to
coordinate their actions.  We use the metavariable $\sigma$ to range over
\emph{sessions}, which comprise the keys of the protocol participants
($\sigma.\init$ and $\sigma.\resp$) of each participant as well as their nonces
($\sigma.\INITSHARE$ and $\sigma.\RESPSHARE$).  The key of the session is
defined as
$\sigma.\KEY \eqdef \key{\SENC}{[\pkey{\sigma.\init}; \pkey{\sigma.\resp};
  \sigma.\INITSHARE; \sigma.\RESPSHARE]}$.

\def\NSL{\mathsf{NSL}}
\newcommand{\SESSIONNSL}{\session_{\NSL}}
\newcommand{\sessionNSL}[3]{\SESSIONNSL\,{#1}\,{#2}\,{#3}}

\begin{theorem}
  \label{thm:nsl-correctness}
  Define
  \begin{align*}
    \sessionNSL{\sk_I}{\sk_R}{\sigma}
    & \eqdef
      \sk_I = \sigma.\init * \sk_R = \sigma.\resp \\
    & \quad {} *
      \always (\public{\sigma.\cf{key}} \iff \later (\public{\sk_I} \lor
      \public{\sk_R}))
  \end{align*}
  The following triples are valid:
  \begin{mathpar}
    \hoareHV
    {
      \TRUE
    }
    { \cf{initiator}\,\sk_I\,(\pkey{\sk_R}) }
    {
      \begin{array}{rl}
        r, & r \neq \None \wand \exists\sigma, r = \Some\,\sigma.\cf{key} \\
        & \quad {} * \sessionNSL{\sk_I}{\sk_R}{\sigma} \\
        & \quad {} * \token{\sigma.\cf{init}}{(\upclose{\nsSess})}
      \end{array}
    }
    \and
    \hoareHV
    { \TRUE }
    { \cf{responder}\,\sk_R }
    {
      \begin{array}{rl}
        r, & r \neq \None \wand \exists\sk_I\,\sigma, \\
           & \quad r = \Some\,(\pkey{\sk_I}, \sigma.\cf{key}) \\
           & \quad {} * \sessionNSL{\sk_I}{\sk_R}{\sigma} \\
           & \quad {} * \token{\sigma.\cf{resp}}{(\upclose{\nsSess})}
      \end{array}
    }
  \end{mathpar}
\end{theorem}

Let us dissect this result.  We focus on the initiator, since the responder is
similar.
If the protocol successfully terminates, the function returns the session key
exchanged by the two agents.  The predicate $\sessionNSL{\sk_I}{\sk_R}{\sigma}$
says that the session key $\sigma.\KEY$ is public if and only if one of the
long-term secret keys is known by the attacker.

\begin{figure}[t]
  \centering
  \small
  \begin{align*}
    I_{\cf{m1}}(\sk_R, m_1)
    & \eqdef \exists\,a\,\sk_I, m_1 = [a; \pkey{\sk_I}]
    * (\public{a} \iff \later (\public{\sk_I} \lor \public{\sk_R})) \\
    I_{\cf{m2}}(\sk_I,m_2)
    & \eqdef \exists\,a\,b\,\sk_R, m_2 = [a;b;\pkey{\sk_R}]
      * (\public{b} \iff \later (\public{\sk_I} \lor \public{\sk_R})) \\
    I_{\cf{m3}}(\sk_R,m_3)
    & \eqdef \TRUE
  \end{align*}
  \caption{Invariants used in NSL proof}
  \Description{Invariants used in NSL proof}
  \label{fig:nsl-invariants}
\end{figure}

To prove the specifications, we use the message invariants of
\Cref{fig:nsl-invariants}. Each invariant conveys to the recipient of a message
what they need to know to conclude that the postcondition holds.  (Note that the
third invariant is trivial.  We will come back to this point later.)
Consider how we prove the correctness of the initiator.  We use the {\sc
  MkNonce} rule to generate a fresh nonce $a$ such that
$\public{a} \iff \later(\public{\sk_I} \lor \public{\sk_R})$.  This nonce comes
with a token resource, which we will use in the postcondition.
To send $m_1$ to the responder, we need to show that it is public, which boils
down to (1) proving $I_{\cf{m1}}$ and (2) proving
$\public{\sk_R} \Rightarrow \public{a} * \public{\pk_I}$.  Both points follow
from how $a$ was generated.
Now, consider what happens when the initiator receives $m_2$. Since the message
is public, after decrypting and checking it, we need to consider two cases
(cf. \Cref{fig:cryptis}).  One possibility is that the body of the encrypted
message (that is, the nonces $a$ and $b$) is public, which could happen if $m_2$
was sent by a malicious party.  Because $b$ is public, the third message $m_3$
is safe to send.  Moreover, because $a$ turned out to be public, it must be the
case that either $\sk_I$ or $\sk_R$ is compromised.  Since $\public{a}$,
$\public{b}$ and $\public{\sk_I} \lor \public{\sk_R}$ hold, we can trivially
prove the logical equivalence in $\SESSIONNSL$ is valid.

The other possibility is that the invariant of $m_2$ holds.  The secrecy
assertion on $b$ allows us to prove $\public{\sk_R} \Rightarrow \public{b}$,
thus the third message is safe to send. To conclude, we need to prove that the
session key $k$ has the desired secrecy.  This follows because the invariant on
$m_2$ guarantees that
$\public{b} \iff \later (\public{\sk_I} \lor \public{\sk_R})$ holds, and because
of how $a$ was generated.

\subsection{What about the third message?}

Our proof of NSL did not require any particular invariants of the third message.
In fact, whether the third message is needed or not depends on what the protocol
is being used for.

The original purpose of the NSL protocol, as devised by its
authors~\cite{NeedhamS78}, is to establish an authenticated interactive session
between two parties. Indeed, this seems to be the purpose of most authentication
protocols, which usually involve the exchange of a session key.  In this case,
the third message is redundant, because it does not add anything on top of the
session messages exchanged later.  All that we need to know is that the session
key is known only to the relevant parties and suitably fresh to prevent replay
attacks.  As we will see, metadata tokens provide a mechanism to argue about
freshness: since we cannot own a token that overlaps with existing metadata
(cf. \Cref{fig:cryptis}), we can guarantee that a fresh key $k$ does not belong
to a set of old keys $K$ by ensuring that all keys in $K$ are generated from
sub-terms whose tokens have been consumed.

However, we could consider alternative scenarios where the protocol
authenticates a \emph{single} request or message.  For example, in a protocol
for financial transactions, the client might not need to establish a whole
interactive session with the server just to send one request.  We could embed
the request data $d$ in the handshake messages, and the last message would serve
as a confirmation step to transfer funds to a vendor or carry out whatever other
action is requested.

We could adapt our proof to reason about this pattern as follows.  When the
server allocates the nonce $b$, it uses its metadata to record the client data
$d$ and includes this metadata assertion with the server's reply.  We can then
include this assertion in the invariant of the third message, along with an
escrow~\cite{KaiserDDLV17} that allows the server to trade in one of $b$'s
tokens against some resource $P~d$ that is needed to process the request.  This
idea would also make sense in hybrid scenarios, such as in the early data
extension of TLS 1.3, where the handshake messages can be used to carry some
application-level data before the regular message exchange begins.  (In
\Cref{sec:reliable-connections}, we will discuss in more detail how escrows in
Cryptis enable the transfer of resources through messages.)

The guarantees of the third message are also tied to another point that we
discussed earlier: the temporal aspect of authentication.  In most tools for
protocol verification, the specification of an authentication protocol includes
trace properties stating that various belief events logged by the agents occur
in a certain order.  If we were interested in adding such temporal guarantees to
\Cref{thm:nsl-correctness}, we might need to strengthen the third invariant to
communicate this event to the responder.  Our results, however, do not involve
such temporal properties because they do not play a role when reusing the
specifications to prove other types of results.  In DY*~\cite{Bhargavan21}, for
example, we can attach invariants to specific protocol events, similarly to how
encrypted messages carry their own invariants.  However, because events are
ghost code, the only way we can learn that an event occurred is by triggering it
or by inspecting the invariant of another event or message.  Thus, it would be
possible to do away with event invariants in DY* and use message invariants to
record the same properties.  In other words, the presence of protocol events is
only useful to reason about protocol events.  We can compare this situation with
the specification of a function that manipulates an imperative data structure.
We could, in principle, consider a semantics with a global trace of allocations,
which would allow us to reason about the exact order in which different parts of
the data structure were created.  However, most verification tools do not bother
doing so.

\subsection{Game Security for NSL}

\begin{figure}[t]
  \centering
\small
\begin{minipage}[t]{0.5\linewidth}
\begin{BVerbatim}
let check_key_secrecy session_key =
  let guess = recv () in
  assert (session_key != guess)

(* keysI: Set of agreed session keys  *)
(* skI: Honest initiator's secret key *)
(* pkR: Honest responder's public key *)

let rec do_init keysI skI pkR =
  fork (fun () -> do_init keysI skI pkR);
  (* Attacker chooses responder *)
  let pkR' = recv () in
  (* Run handshake *)
  let k = init skI pkR' in
  (* The session key should be fresh and *)
  assert (not (Set.mem keysI k));
  Set.add keysI k;
  (* if attacker chose honest responder,
     the key cannot be guessed. *)
  if pkR' == pkR then check_key_secrecy k
  else ()
\end{BVerbatim}
\end{minipage}%
\begin{minipage}[t]{0.5\linewidth}
\begin{BVerbatim}
let rec do_resp keysR skR pkI =
  (* Similar to initiator *)
  (* ... *)

let game () =
  (* Generate keys and leak public keys *)
  let skI, skR =
    mk_aenc_key (), mk_aenc_key () in
  let pkI, pkR = pkey skI, pkey skR in
  send pkI; send pkR;
  (* Generate sets of session keys *)
  let keysI = Set.new () in
  let keysR = Set.new () in
  (* Run agents *)
  fork (fun () -> do_init keysI skI pkR);
  fork (fun () -> do_resp keysR skR pkI)

\end{BVerbatim}
\end{minipage}%
\caption{A security game where the attacker tries to learn the session keys or
  cause them to be reused.}
  \Description{Implementation of the NSL protocol}
  \label{fig:nsl-game}
\end{figure}

Because \Cref{thm:nsl-correctness} does not involve the traditional temporal
properties used in protocol verification, we might worry that it might be
leaving the door open to an attack.
To increase our confidence in this result, we use a symbolic security game
(\Cref{fig:nsl-game}). We generate fresh keys for two honest participants, an
initiator and a responder, and let them run an arbitrary number of NSL sessions
in parallel (\verb!do_init! and \verb!do_resp!).  In each iteration of
\verb!do_init!, the initiator attempts to contact an agent chosen by the
attacker. If the handshake successfully terminates, the initiator adds the
exchanged session key to a set of keys \verb!keysI!, while ensuring that the key
is fresh.  Moreover, if the initiator contacted the honest responder, the
attacker tries to guess the session key.  The session is successful if its key
had not been previously used and cannot be guessed by the attacker.  The logic
in \verb!do_resp! is similar.  This implies that the assertion in
$\cf{check\_key\_secrecy}$ cannot fail, because terms that come from the network
are public, and because the attacker does not know $\sk_I$ or $\sk_R$.

Providing this kind of guarantee can be elusive.  Indeed, the original version
of the NSL protocol~\cite{NeedhamS78} was vulnerable to a man-in-the-middle
attack~\cite{Lowe96}, even though it was thought to be secure for several years
(and even verified~\cite{BurrowsAN90}). The issue was that the original version
omitted the identity of the responder in $m_2$---that is, $m_2$ would have been
$\aenc{\pk_I}{[\tf{m2};a;b]}$ instead of $\aenc{\pk_I}{[\tf{m2};a;b;\pk_R]}$.
This omission meant that the initiator had no way of telling if the responder
was actually supposed to see the nonce $b$.  Indeed, the second invariant ties
the confidentiality of $b$ to the secret key of the responder, and this property
is required to prove that the third message can be sent.  If the responder's
identity were not explicitly mentioned in the message, it would be impossible to
know who can see $b$, so it would be impossible to prove that the third message
is safe.

As seen in \Cref{fig:lowe's-attack}, a malicious responder $M$ can exploit this
issue to lead an honest $R$ into generating a nonce $b$ for authenticating $I$,
and then tricking $I$ into leaking this nonce to $M$.  In the end, $M$ is able
to construct the same session key that $R$ believes is being used to talk to
$I$---despite the fact that $R$ believes that the handshake was performed
between two agents that are, in fact, honest.  The game shows that the attack
cannot succeed---otherwise, $\cf{check\_key\_secrecy}$ would fail.

\begin{figure}[t]
  \centering
  \small
\begin{minipage}{0.5\linewidth}
  \begin{align*}
    I \to M & : \aenc{\pk_M}{[\cf{m1}; a; \pk_I]} \\
    M \to R & : \aenc{\pk_R}{[\cf{m1}; a; \pk_I]} \\
    R \to M & : \aenc{\pk_I}{[\cf{m2}; a; b]}
  \end{align*}
\end{minipage}%
\begin{minipage}{0.5\linewidth}
  \begin{align*}
    M \to I & : \aenc{\pk_I}{[\cf{m2}; a; b]} \\
    I \to M & : \aenc{\pk_M}{[\cf{m3}; b]} \\
    M \to R & : \aenc{\pk_R}{[\cf{m3}; b]}.
  \end{align*}
\end{minipage}%
\caption{Attack on the original Needham-Schroeder protocol~\cite{Lowe96}.}
  \label{fig:lowe's-attack}
\end{figure}

To show that the attacker cannot win the game, we proceed as follows.  First, we
prove specifications for the functions \verb!do_init! and \verb!do_resp!  that
guarantee that they are safe to run.  We consume the secrecy resources of the
agents private keys to guarantee that they cannot become public.  In the proof
of \verb!do_init!, we invoke the specification of \verb!init! in
\Cref{thm:nsl-correctness}.  We maintain an invariant on \verb!keysI!
guaranteeing that every key $k'$ stored in the set satisfies
$\meta{k'}{\nsSess}{()}$.  This means that the new session key $k$ cannot be in
the set, because its token has not been used yet.  Thus, the first assertion
cannot fail. We consume this token so that the key can be added to \verb!keysI!.
We then argue that the second assertion cannot fail because the attacker's guess
is public, whereas the session key cannot be session is authentic.
A symmetric reasoning shows that \verb!do_resp! is safe as well.
Finally, we prove that \verb!game! is safe.  We generate the key pairs of the
honest participants by invoking the specifications in \Cref{sec:cryptis}.  Then,
we allocate two empty sets of keys, which trivially satisfy the invariant that
all keys have their metadata token set.  We conclude by invoking the
specifications of \verb!do_init! and \verb!do_resp! to show that the last line
is safe.

\aaa{We could also have a game where the $I$ and $R$ run a single handshake, and
  where we check that the responder contacted by $I$ is $R$ if and only if the
  initiator that contacts $R$ is $I$.  This would involve strengthening the
  invariant of the third message to include a resource that would allow us to
  reason about this property.  But this point is subtle and confusing, and I
  don't know if it adds that much to the paper.}

\section{Diffie-Hellman key exchange and forward secrecy}
\label{sec:iso}

\begin{figure}
\centering
\small
\begin{minipage}[t]{0.5\linewidth}
\begin{BVerbatim}
let initiator skI pkR =
  let pkI = pkey skI in
  let a = mknonce () in
  let m1 = ["m1"; g^a; pkR] in
  send m1;
  let m2 = recv () in
  let [="m2"; =g^a; gb; =pkI] =
    verify pkR m2 in
  let m3 = sign skI ["m3"; g^a; gb; pkR] in
  send m3;
  let k =
    key senc [pkI; pkR; g^a; gb; gb^a] in
  k

\end{BVerbatim}
\end{minipage}%
\begin{minipage}[t]{0.5\linewidth}%
\begin{BVerbatim}
let responder skR =
  let pkR = pkey skR in
  let m1 = recv () in
  let [="m1"; ga; pkI] = m1 in
  let b = mknonce () in
  let m2 = sign skR ["m2"; ga; g^b; pkI] in
  send m2;
  let m3 = recv () in
  let [="m3"; =ga; =g^b; =pkR] =
    verify pkI m3 in
  let k =
    key senc [pkI; pkR; ga; g^b; ga^b] in
  (pkI, k)

\end{BVerbatim}
\end{minipage}
\caption{ISO authentication protocol based on Diffie-Hellman key exchange.  }
\Description{Authentication based on Diffie-Hellman key exchange}
  \label{fig:iso-dh}
\end{figure}

One limitation of a protocol such as NSL is that it is vulnerable to \emph{key
  compromise}.  If a private key is leaked, an attacker can decrypt the messages
of a handshake and learn its session key.  By contrast, many modern protocols
guarantee \emph{forward secrecy}: if a handshake is successful, its session keys
will remain secret even if long-term keys are leaked~\cite{Cohn-GordonCG16}.
Our goal in this section is to demonstrate that Cryptis can scale up to more
complex protocols with richer guarantees.  Specifically, we will prove the
correctness of the ISO protocol~\cite{Krawczyk03}, which provides forward
secrecy.  Because of its stronger guarantees, it will be our protocol of choice
to implement the communication components used in our key-value store.  A
typical run of the protocol proceeds as follows:
\begin{align*}
  I \to R & : [\tf{m1}; g^a; \pk_I] &
  R \to I & : \sign{\sk_R}{[\tf{m2}; g^a; g^b;\pk_I]} &
  I \to R & : \sign{\sk_I}{[\tf{m3}; g^a; g^b; \pk_R]}.
\end{align*}
The flow is similar to the NSL protocol, except that (1) it uses digital
signatures instead of asymmetric encryption; (2) the first message does not need
to be signed or encrypted; (3) the keys used in the signed messages 2 and 3 are
swapped; (4) the agents exchange the key shares $g^a$ and $g^b$ rather than the
nonces $a$ and $b$. At the end of the handshake, the participants can compute
the shared secret $g^{ab} = (g^a)^b = (g^b)^a$ and use it to derive a session
key. \Cref{fig:iso-dh} shows an implementation of ISO.

We state security following the idea of \Cref{sec:nsl}. We formulate a
specification for the initiator and the responder, and use these specifications
to prove the security of a game between the attacker and the agents.  The main
difference lies in the secrecy guarantees for the session key $k$: when the
handshake terminates, if we can prove that the participants' long-term keys are
not compromised \emph{yet}, then $k$ will remain secret forever, even if some
long-term keys are leaked later.  The ISO session $\sigma$ now includes a
component $\sigma.\SECRET$, which corresponds to the shared Diffie-Hellman
secret.  We define $\sigma.\KEY$ as
$\key{\SENC}{[\sigma.\init; \sigma.\resp; \sigma.\INITSHARE; \sigma.\RESPSHARE;
  \sigma.\SECRET]}$.

\newcommand*{\SESSIONISO}{\session_{\cf{ISO}}}
\newcommand*{\sessionISO}[3]{\SESSIONISO~{#1}~{#2}~{#3}}
\newcommand*{\RELEASED}{\cf{released}}
\newcommand*{\released}[1]{\RELEASED\,#1}
\newcommand*{\RELEASETOKEN}{\cf{release\_token}}
\newcommand*{\releaseToken}[1]{\RELEASETOKEN\,#1}
\newcommand*{\nsRel}{\mathtt{ISO.released}}
\newcommand*{\nsISO}{\mathtt{ISO}}

\begin{theorem}
  \label{thm:iso-dh-correctness}
  Define
  \begin{align*}
    \sessionISO{\sk_I}{\sk_R}{\sigma}
    & \eqdef \sk_I = \sigma.\init * \sk_R = \sigma.\resp \\
    & \quad {} * (\public{\sk_I} \lor \public{\sk_R} \lor
      \always (\public{\sigma.\cf{key}} \iff \later \FALSE)).
  \end{align*}

  The following triples are valid:
  \begin{mathpar}
    \hoareV
    { \TRUE }
    { \cf{initiator}\,\sk_I\,\pk_R }
    {
      \begin{array}{rl}
        r, & r = \None \lor \exists\,\sk_R\,\pk_R\,\sigma, \\
           & \quad {} r = \Some\,k * \signKey{\sk_R}{\pk_R} \\
           & \quad {} * \sessionISO{\sk_I}{\sk_R}{\sigma} \\
           & \quad {} * \token{\sigma.\cf{init}}{(\top \setminus \upclose{\cf{ISO}})}
      \end{array}
    }
    \and
    \hoareV
    { \TRUE }
    { \cf{responder}\,\sk_R }
    {
      \begin{array}{rl}
        r, & r = \None \lor \exists\,\sk_I\,\pk_I\,\sigma, \\
           & \quad {} r = \Some\,(\pk_I, k) * \signKey{\sk_I}{\pk_I}\\
           & \quad {} * \sessionISO{\sk_I}{\sk_R}{\sigma} \\
           & \quad {} * \token{\sigma.\cf{resp}}{(\top \setminus \upclose{\cf{ISO}})}
      \end{array}
    }
  \end{mathpar}
\end{theorem}

\begin{figure}
  \centering
  \small
  \begin{align*}
    I_{\cf{m2}}(\sk_R,m_2)
    & \eqdef \exists s_a\,b\,\pk_I,  m_2 = [s_a; g^b; \pk_I] * (\public{b} \iff \later \FALSE) \\
    I_{\cf{m3}}(\sk_I, m_3)
    & \eqdef \exists a\,s_b\,\sk_R, m_3 = [g^a; s_b; \pkey{\sk_R}]
      * (\public{\sk_I} \lor \public{\sk_R} \lor \\
    & \qquad
      (\public{(\key{\SENC}{[\pkey{\sk_I}; \pkey{\sk_R}; g^a;
      s_b; s_b^a]})} \wand \later \FALSE))
  \end{align*}
  \caption{Invariants for ISO protocol.}
  \Description{Invariants for ISO protocol.}
  \label{fig:iso-dh-invariants}
\end{figure}

We use the invariants of \Cref{fig:iso-dh-invariants}.  Each agent allocates
their nonces $n$ so that $\public{n} \iff \later \FALSE$.  When $I$ checks the
signature, either $R$'s secret key is compromised, or they learn that $R$'s key
share is of the form $g^b$, with $\public{b} \iff \later \FALSE$.  Since
$\public{a} \iff \later \FALSE$, \Cref{fig:cryptis} implies that
$\public{g^{ab}}$ is equivalent to $\later \FALSE$.
We modify the game of \Cref{fig:nsl-game} so that both long-term secret keys are
eventually leaked, and we only check a session key if it was exchanged before
the compromise (\Cref{fig:iso-dh-game}).
To prove that the game is secure, we proceed similarly to what we did for the
NSL game.  The main difference lies in the management of long-term keys.  After
generating the signature keys $\sk_I$ and $\sk_R$, we allocate an invariant $I$
that says that either the compromise bit $c$ is set to false, in which case
$\secret{\sk_I} * \secret{\sk_R}$ holds, or it is set to true, in which case
both $\sk_I$ and $\sk_R$ are public.
Then, we prove that the \verb!check_key_secrecy! function is safe provided that
it is called on a session key $k$ of the ISO protocol.  If we run the ``then''
branch of that function, the invariant $I$, combined with the postcondition of
the handshake, implies that $\always (\public{k} \iff \later \FALSE)$ holds.  By
a reasoning analogous to the NSL game, this guarantees that the attacker cannot
win.

\begin{figure}[t]
  \centering
\small
\begin{minipage}[t]{0.5\linewidth}
\begin{BVerbatim}
(* c: Have keys been compromised? *)
let rec wait_for_compromise c =
  if !c then () else wait_for_compromise c

let check_key_secrecy c k =
  if not !c then
    wait_for_compromise c;
    let guess = recv () in
    assert (k != guess)
  else ()

let compromise_keys c skI skR =
  c := true; send skI; send skR
\end{BVerbatim}
\end{minipage}%
\begin{minipage}[t]{0.5\linewidth}
\begin{BVerbatim}
let game () =
  (* ... *)
  let skI = mk_sign_key #() in
  let skR = mk_sign_key #() in
  let pkI = pkey skI in
  let pkR = pkey skR in
  let c = ref false in
  (* ... *)
  fork (fun () -> do_init keysI c skI pkR);
  fork (fun () -> do_resp keysR c skR pkI);
  fork (fun () -> compromise_keys c skI skR)

\end{BVerbatim}
\end{minipage}%
\caption{Security game for the ISO protocol (excerpt).  The agents' secret
  signing keys are leaked eventually, and we only check the secrecy of session
  keys if they have been exchanged before the compromise.}
  \Description{Implementation of the NSL protocol}
  \label{fig:iso-dh-game}
\end{figure}

\paragraph*{Decomposing the responder}

In a typical client/server setting, it is useful to decompose $R$'s logic into
two steps.  In the \verb!ISO.listen! function, the responder waits for an
incoming connection request, the first message of the ISO protocol.  The server
can use the initiator's identity to decide whether to accept the connection or
not.  If it decides to accept the connection, it can call the \verb!ISO.confirm!
function, which generates the responder's key share and runs the rest of the ISO
handshake.

\paragraph*{Session compromise}

The specification of ISO has a limitation: if the handshake completes
successfully, it becomes impossible for us to model the compromise of the
session key $k$.  We can relax this limitation by modifying the secrecy
predicates of the private DH keys $a$ and $b$ so that
\[ \public{a} \iff \public{b} \iff \later (\released{g^a} * \released{g^b}), \]
where the $\RELEASED$ predicate is just a wrapper around term metadata that
obeys the following rules:
\begin{mathpar}
  \releaseToken{t} * \released{t} \wand \FALSE

  \and

  \releaseToken{t} \vs \released{t}
\end{mathpar}
We strengthen the ISO invariants and \Cref{thm:iso-dh-correctness} so that the
postconditions include a release token resource for the DH public key of the
corresponding agent.  Then, we can model a compromise of the session key by
simply releasing the tokens of the initiator and the responder.  While the
agents still hold their release tokens, we can prove that the session key is not
yet compromised.

\paragraph*{Early compromise}

Conversely, if we know that one of the agents is already compromised before
initiating a handshake, it is useful to have the session key $k$ be public from
the start.  In this way, if we use the session key to encrypt something
(cf. \Cref{sec:reliable-connections}), we do not need to prove the corresponding
message invariants.  Our complete specifications of the ISO protocol allow us to
choose $k$ to be public in this scenario.  To accommodate this, we add a new
parameter $\rho$ to the session predicate, which tracks whether the agent of
role $\rho \in \{\init, \resp\}$ was able to compromise the session early.

\section{Reliable Connections}
\label{sec:reliable-connections}

\aaa{We need to explain how this resource transfer compares to
  \textcite{GondelmanHPTB23}}

Now that we have an authentication protocol, we can use it to implement
authenticated, reliable connections.  At the logic level, we follow prior work
and model this functionality as the ability to reliably transfer arbitrary
separation-logic resources~\cite{HinrichsenBK20,GondelmanHPTB23}.
Operationally, the functionality guarantees that messages are received in the
same order that they are sent and that their contents are not modified.  To
preserve their order, we include sequence numbers in every message sent; to
preserve their contents, we encrypt them with a session key.
\begin{figure}[t]
  \centering
\small
\begin{minipage}[t]{0.45\linewidth}
\vspace{0pt}
\begin{BVerbatim}
let Conn.listen () = ISO.listen ()

let Conn.confirm skR request =
  let k = ISO.confirm skR request in
  {session_key = k;
   sent = 0; received = 0}

let Conn.send conn s m =
  let ciphertext =
    senc conn.session_key
      [s; conn.sent; m] in
  conn.sent++;
  send ciphertext
\end{BVerbatim}
\end{minipage}%
\begin{minipage}[t]{0.55\linewidth}%
\vspace{0pt}
\begin{BVerbatim}
let Conn.connect skI pkR =
  let k = ISO.initiator skI pkR in
  {session_key = k;
   sent = 0; received = 0}

let Conn.recv conn s =
  let rec loop () =
    let m = recv () in
    let [=s; n; payload] = 
      sdec c.session_key m in
    if n == c.received then
      c.received++; payload
    else loop ()
  in loop ()
\end{BVerbatim}
\end{minipage}
\vspace{2em}
\begin{mathpar}
  \inferrule[ConnConnect]
  { }
  {
    \hoare
    { \public{\sk_C} \lor \public{\sk_S} \lor P }
    { \cf{Conn.connect}~\sk_C~\pk_S }
    { c,
      \setlength{\arraycolsep}{0pt}
      \begin{array}{l}
        \cConnConnected~\cf{init}~c \\
        {} * (\COMPROMISED~\cf{init}~c \lor P) \\
        {} * \releaseToken{c.\cf{init}} \\
        {} * \token{c.\cf{init}}{(\top \setminus \upclose{\cf{ISO}}
        \setminus \upclose{\cf{Conn}})}
      \end{array}
    }
  }

  \inferrule[ConnConfirm]
  { }
  {
    \hoare
    { \setlength{\arraycolsep}{0pt}
      \begin{array}{l}
        \public{\vf{ga}} \\
        {} * (\public{\sk_C} \lor \public{\sk_S} \lor P)
      \end{array}
    }
    { \cf{Conn.confirm}~\sk_S~(\vf{ga}, \pk_C) }
    { c,
      \setlength{\arraycolsep}{0pt}
      \begin{array}{l}
        \cConnConnected~\cf{resp}~c \\
        {} * (\COMPROMISED~\cf{resp}~c \lor P) \\
        {} * \releaseToken{c.\cf{resp}} \\
        {} * \token{c.\cf{resp}}{(\top \setminus \upclose{\cf{ISO}}
        \setminus \upclose{\cf{Conn}})}
      \end{array}
    }
  }

  \inferrule[ConnSend]
  { \meta{\SENC}{\namesp}{\connPred{\rho}{\varphi}} }
  {
    \hoare
    { \cConnConnected~\rho~c * \public{\vec{t}} * (\public{c.\cf{key}} \lor \varphi~\sk_C~\sk_S~c~\vec{t})}
    { \cf{Conn.send}~{c}~\namesp~\vec{t} }
    { \cConnConnected~\rho~c }
  }

  \and

  \inferrule[ConnRecv]
  { \meta{\SENC}{\namesp}{\connPred{\rho^{-1}}{\varphi}} }
  { \hoare
    { \cConnConnected~\rho~c }
    { \cf{Conn.recv}~c~\namesp }
    { \vec{t}, \cConnConnected~\rho~c * \public{\vec{t}} * (\public{c.\cf{key}} \lor
      \varphi~\sk_C~\sk_S~c~\vec{t})
    }
  }

  \and

  \COMPROMISED~\rho~c \vdash \public{c.\cf{key}}

  \and

  \cConnConnected~\rho~c *
  \releaseToken{c.\rho} *
  \public{c.\cf{key}}
  \vdash \later \COMPROMISED~\rho~c

  \and

  \secret{\sk_C} *
  \secret{\sk_S} *
  \cConnConnected~\rho~c
  \vdash \later \always \neg \COMPROMISED~\rho~c
\end{mathpar}
\caption{Implementation and specification of reliable communication.  The
  variable $\rho \in \{ \cf{init}, \cf{resp} \}$ denotes the role of an agent,
  and $\rho^{-1}$ denotes the opposite role.}
\label{fig:reliable-connections}
\end{figure}

The implementation and the specifications of this functionality are described in
\Cref{fig:reliable-connections}.  By abuse of notation, we sometimes use a
connection object $c$ as if it were is underlying session $\sigma$.  In
particular, $c.\rho$ refers to the key share of the agent of role $\rho$.  There
is no harm in doing that because the connection object tracks the session key,
which fully determines all the session information.

To connect to a server, a client uses the $\cf{connect}$ function, which
initiates a handshake of the ISO protocol and stores the resulting session key
in the returned connection object, along with counters for tracking sequence
numbers.  The server behaves similarly, but runs the responder of the protocol.
Once a connection is established, we can use the $\send$ and $\recv$ functions
to communicate.  This functions include and check the sequence numbers stored in
the connection object to ensure that messages are received in the appropriate
order.  The $\recv$ function keeps polling the network until it receives a
message with the expected tag and sequence number.

Let us analyze these specifications.  As we mentioned earlier, it will be useful
to let the connection functions create compromised connections if we know that
one of the participants is also compromised.  Accordingly, the precondition of
the functions $\cf{connect}$ and $\cf{confirm}$ assumes that either the agents
have been compromised or some resource $P$ is available.  When the connection is
established, it will be marked as compromised if the first case holds;
otherwise, the resource $P$ will be available for use.  This allows us to
maintain an invariant $P$ across multiple connections, provided that the
protocol participants are not compromised.  Moreover, the postconditions provide
release tokens to compromise session keys and a metadata token.  Finally, the
postconditions provide the resource $\cConnConnected~\sk_C~\sk_S~\rho~c$, which
says that the connection is ready.
To send and receive messages, we must have assigned special
$\connPred\rho\varphi$ predicates to their tags.  For simplicity, we assume that
each tag can be used to send messages for a single role $\rho$.  Then, to send a
message $\vec{t}$, we must prove that $\varphi~\sk_C~\sk_S~c~\vec{t}$ holds,
unless the session key is compromised.  Dually, we can assume that this
invariant holds when receiving the message.  Crucially, these invariants are
\emph{not} required to hold persistently, which allows us to transfer resources
through a connection.

\renewcommand*{\sessionISO}[4]{\SESSIONISO~#1~#2~#3~#4}

\begin{figure}[t]
  \centering
  \small
\begin{align*}
  \cConnConnected~\sk_C~\sk_S~\rho~c
  & \eqdef
    \setlength{\arraycolsep}{0pt}
    \begin{array}[t]{l}
      \sessionISO{\sk_C}{\sk_S}{\rho}{c} \\
      {} * \exists n\,m, c \mapsto \{\vf{key} = k; \vf{sent} = n;
      \vf{received} = m\}  * \termOwn{c.\rho}{\cf{conn}}{\authfull m}
    \end{array} \\
  \connPred{\rho}{\varphi}
  & \eqdef
    \setlength{\arraycolsep}{0pt}
    \begin{array}[t]{l}
      \lambda k\, t, \exists \sigma\,n\,\vec{t},
      k = \sigma.\cf{key} * t = (n :: \vec{t}) * \public{\vec{t}} \\
      {} * (\termOwn{c.\rho^{-1}}{\cf{conn}}{\authfull n}
      \vs[\top] \later (\varphi~\sigma.\init~\sigma.\resp~\sigma~\vec{t} * \termOwn{k}{\cf{conn}.\rho^{-1}}{\authfull
      n+1}))
    \end{array}
\end{align*}
  \caption{Predicates for reliable connections}
  \label{fig:reliable-connections-preds}
\end{figure}

To enable this transfer of resources, we use a variant of the \emph{escrow
  pattern}~\cite{KaiserDDLV17,GondelmanHPTB23}, where a party can extract a
resource $R$ from an invariant $I$ by exchanging it against a guard $P$.  This
can be implemented by having $I$ hold some resource $R'$ such that
$\later (R' * P) \vs \later (R' * R)$; for example, if $P$ is exclusive
($P * P \vdash \FALSE$), we can take $R'$ to be $R \lor P$.
\Cref{fig:reliable-connections-preds} presents the definitions of the predicates
used in the specifications.  The resource $\cConnConnected~\sk_C~\sk_S~\rho~c$
contains an assertion $\termOwn{k.\rho}{\cf{conn}}{\authfull m}$ which tracks
how many messages that agent has received and serves as the guard of the escrow.
The message invariant $\connPred{\rho}{\varphi}$ contains an escrow that allows
us to trade in that guard against resources attached to the message payload,
provided that the guard's counter matches the sequence number of the message.
Note that this escrow also returns an updated guard, signaling the fact that
another message was received.

\aaa{We could expand on this to include key ratcheting; that is, have the agents
  derive new keys while sending the messages of the protocol.  It would be
  technically interesting, but would make the paper even more dense.}

\section{Remote Procedure Calls}
\label{sec:rpc}

\newcommand*{\cRpcConnected}{\cf{RPC.connected}}
\newcommand*{\cRpcServerConnected}{\cf{RPC.Server.connected}}
\newcommand*{\cRpcCall}{\cf{RPC.call}}
\newcommand*{\cRpcClose}{\cf{RPC.close}}
\newcommand*{\cRespPred}{\cf{resp\_pred}}
\newcommand*{\cRespPredToken}{\cf{resp\_pred\_token}}

The last internal component we need for our key-value store is the RPC
mechanism.
The component is a thin layer on top of a reliable communication functionality
(cf. \Cref{fig:reliable-connections}).
The connection stage is mostly unchanged.  When the server accepts a connection
from a client $C$, it enters a loop that continuously receives requests from
$C$.
To perform a call ($\cRpcCall$), the client sends message with the appropriate
tag (which identifies the server operation) and its arguments and then wait for
the corresponding response.
The server invokes a handler on this request based on the operation and sends
whatever the handler returns back to the client.

Each RPC operation comes with two predicates: a predicate $\varphi$ that should
hold of the arguments of the operation, and a predicate $\psi$ for its return
values.  It is convenient for $\psi$ to be able to refer to the arguments in
addition to the results.  Here, the ability to transfer resources with reliable
connections comes in handy.  Using a fractional agreement algebra and term
metadata, we define a resource $\cRespPredToken_q~\sigma~\psi'$, which keeps
track of which property $\psi'$ should hold of the results.  The RPC client
allocates this resource when the connection is established.  Before performing a
call, the client updates $\psi'$ to $\psi~\sk_C~\sk_C~\sigma~\vec{t}$, where
$\vec{t}$ is the list containing the arguments of the operation.  Then, it sends
one half of this token to the server.  The message predicate for the server's
response says that the results of the operation should satisfy exactly this
predicate.

The RPC functionality also includes a close call, which allows the client to
close the connection with the server.  To close the connection, the client
consumes its release token and informs the server that the token has been
released.  The server releases its token as well, at which point the session key
becomes public.  This allows the server to reply to the client without proving
any particular invariants.  When the client receives the server's
acknowledgment, they conclude that the session key has been made public.

\begin{figure}[t]
  \centering
\small
\begin{align*}
  \cRespPredToken_q~\sigma~\varphi
  & \eqdef \termOwn{\sigma.\INITSHARE}{\cf{rpc.resp\_pred}}{\cf{agree}_q~\varphi} \\
  \rpcPred{\varphi}{\psi}
  & \eqdef
    \connPred{\cf{init}}{
    \left(
    \begin{array}{l}
      \lambda \sk_C\,\sk_S\,\sigma\,\vec{t}, \\
      \cRespPredToken_{1/2}~\sigma~(\psi~\sk_C~\sk_S~\sigma~\vec{t}) \\
      {} * \varphi~\sk_C~\sk_S~\sigma~\vec{t}
    \end{array}\right)} \\
  \cRespPred
  & \eqdef
    \connPred{\cf{resp}}{(\lambda \sk_C\,\sk_S\,\sigma\,\vec{t},
      \exists \psi, \cRespPredToken_{1/2}~\sigma~\psi * \psi \vec{t})} \\
  \cRpcConnected~\sk_C~\sk_S~c
  & \eqdef
    \setlength{\arraycolsep}{0pt}
    \begin{array}[t]{l}
      \cConnConnected~\sk_C~\sk_S~\init~c  * \releaseToken{c.\INITSHARE} \\
      {} * (\COMPROMISED~\sk_C~\sk_S~\init~c \lor \cRespPredToken_1~\sigma~(\lambda \_,
      \FALSE))
    \end{array}
\end{align*}
\vspace{2em}
\begin{mathpar}
  \inferrule[RpcCall]
  { \meta{\SENC}{\namesp}{\rpcPred{\varphi}{\psi}}
    \and \meta{\SENC}{\cf{rpc.resp}}{\cf{resp\_pred}} }
  {
    \hoareV
    { \public{\vec{t}} *
      \cRpcConnected~\sk_C~\sk_S~c *
      (\COMPROMISED~\sk_C~\sk_S~\init~c \lor \varphi~\sk_C~\sk_S~c~\vec{t}) }
    { \cRpcCall~c~\namesp~\vec{t} }
    { \vec{t}', \public{\vec{t}'} *
      \cRpcConnected~\sk_C~\sk_S~c *
      (\COMPROMISED~\sk_C~\sk_S~\init~c \lor \psi~\sk_C~\sk_S~c~\vec{t}~\vec{t'}) }
  }
  \and
  \inferrule[RpcClose]
  { }
  { \hoare
    { \cRpcConnected~\sk_C~\sk_S~c }
    { \cRpcClose~c }
    { \public{c.\KEY} } }
\end{mathpar}
\caption{Remote procedure calls}
  \label{fig:rpc}
\end{figure}

\section{Implementing and verifying the key-value store}
\label{sec:key-value}

\begin{figure}[t]
  \begin{minipage}[t]{0.6\linewidth}%
  \vspace{0pt}
  \centering
  \small
  \caption{Message invariants for storing and loading values.}
  \label{fig:key-value-invariants}
  \end{minipage}%
  \hfill
\end{figure}

\aaa{Add something here about the other separation logic tools}

With all the communication primitives and data structures in place, implementing
the key-value store is relatively straightforward.  The easiest part is
implementing the client wrappers: they are just calls to the RPC mechanism.
The server implementation is more interesting.  We assume that we have an
imperative map module that associates terms with arbitrary values.  Our
implementation uses an association list for simplicity, but we could easily swap
that out for a more efficient implementation.
The server handles the data of several clients simultaneously, with the data of
each client being stored in a separated data structure in memory.  Another data
structure maps client identities (given by their public keys) to their accounts.
An account contains a reference to the client's database and a lock to guarantee
mutual exclusion.
The server runs one thread in a loop waiting for incoming connection requests
from the RPC module.  When a request arrives, the server checks that the
client's account exists and tries to acquire its lock.  If the account doesn't
exist yet, a new one is created.  After the server finds the client's account,
it spawns a new thread to handle incoming RPCs.  Each RPC handler simply
performs the corresponding database operation using the functions of the map
module.  Once the client disconnects, the server releases the account lock and
the handler thread terminates.

\begin{figure}[t]
  \centering
  \small
  \begin{align*}
    & \text{\DbMainAlloc{}}
    & \token{\sk_C}{(\upclose{\namesp_{\cdb}.\client.\sk_S})}
      & \vs \dbState~\emptyset * \top \mapsto_{\cdb} \bot *
        \dbMain~\emptyset * \dbSync~\emptyset \\
    & \text{\DbCopyAlloc{}}
    & \token{\sk_S}{(\upclose{\namesp_{\cdb}.\server.\sk_C})}
      & \vs \dbCopy~\emptyset \\
    & \text{\DbStateAgree{}}
    & \dbState~\delta * t_1 \mapsto_{\cdb} \vf{ot}_2 & \wand \delta~t_1 = \vf{ot}_2 \\
    & \text{\DbStateUpdate{}}
    & \dbState~\delta * t_1 \mapsto_{\cdb} \vf{ot}_2 & \vs \dbState~(\delta[t_1 \mapsto t_2'])
      * t_1 \mapsto_{\cdb} t_2' \\
    & \text{\DbMainUpdate{}}
    & \dbMain~\delta * \dbSync~\delta & \vs \dbMain~\delta' * \dbUpdate~\delta~\delta' \\
    & \text{\DbCopyUpdate{}}
    & \dbCopy~\delta_1 * \dbUpdate~\delta_2~\delta' & \vs \delta_1 = \delta_2
      * \dbCopy~\delta' * \dbSync~\delta' \\
    & \text{\DbMainSync{}}
    & \dbMain~\delta_1 * \dbSync~\delta_2 & \wand \delta_1 = \delta_2
  \end{align*}
  \vspace{0.5em}
  \begin{align*}
    \cDbConnected~\sk_C~\sk_S~c
    & \eqdef \exists\,\delta,
    \cRpcConnected~\sk_C~\sk_S~c \\
    & \qquad {} * \dbState~\delta *
    (\COMPROMISED~\sk_C~\sk_S~c \lor \dbMain~\delta * \dbSync~\delta) \\
    \cDbDisconnected~\sk_C~\sk_S
    & \eqdef \exists\,\delta,
      \dbState~\delta *
    (\public{\sk_C} \lor \public{\sk_S} \lor \dbMain~\delta * \dbSync~\delta)
  \end{align*}
  \vspace{0.5em}
  \begin{align*}
    I_{\cf{store}}~\sk_C~\sk_S~c~\vec{t}
    & \eqdef \exists t_1\,t_2\,\sigma,
      \vec{t} = [t_1, t_2] * \dbUpdate~k~\sigma~\sigma[t_1 \mapsto t_2] \\
    I_{\cf{ack\_store}}~\sk_C~\sk_S~c~\vec{t}~\vec{t}'
    & \eqdef \exists\sigma,\dbSync~k~\sigma \\
    I_{\cf{load}}~\sk_C~\sk_S~c~\vec{t}
    & \eqdef \exists t_1\,t_2\,\sigma,
      \vec{t} = [t_1] * \sigma~t_1 = \Some~t_2 *
      \dbUpdate~\sigma~\sigma
    \\
    I_{\cf{ack\_load}}~\sk_C~\sk_S~c~\vec{t}~\vec{t}'
    & \eqdef \exists t_1\,t_2\,\sigma, \vec{t} = [t_1] * \vec{t}' = [t_2] *
      \sigma~t_1 = \Some~t_2 * \dbSync~\sigma.
  \end{align*}
  \caption{Key-value store: Auxiliary assertions and rules.}
  \label{fig:key-value-assertions}
\end{figure}

To verify the specifications of \Cref{fig:key-value-spec}, we use some custom
resources and RPC predicates described in \Cref{fig:key-value-assertions}.
(Once again, most of these resources are parameterized by the keys of the client
and the server, but we elide most of these parameters for readability.)  We
distinguish between two types of databases: the \emph{logical} database, which
the client believes ought to be stored in the server, and the \emph{physical}
database, which is what is actually stored in the server. The logical database
is ghost state that is owned by the client, and the physical database is tracked
by the resource $\isMap~\vdb~\delta$, which says that the location $\vdb$ points
to an object that represents the map $\delta$.

The logical database consists of a series of resources defined with term
metadata, which the client and the server can initialize by consuming the
appropriate tokens (cf. \DbMainAlloc{} and \DbCopyAlloc{}).  The resource
$\dbState~\sigma$ means that the current logical state is exactly $\sigma$.
This resource behaves similarly to how the heap is modeled in Iris: as shown in
\Cref{fig:key-value-assertions}, it can be combined with the points-to assertion
$t_1 \mapsto_{\cdb} t_2$ to update the logical state (\DbStateUpdate{}) or find
out which values are stored under it (\DbStateAgree{}).

The remaining database resources are ``bridge'' resources that allow us to keep
the logical and the physical databases synchronized.  The resource
$\dbMain~\sigma$ tracks the client's view on the logical database, and
$\dbCopy~\sigma$ tracks the server's view of the physical database.  The
resource $\dbSync~\sigma$ indicates that these two views are in sync.  When the
client wants to update the logical database to $\sigma'$, they consume this
resource to create a new resource $\dbUpdate~\sigma~\sigma'$ with the
\DbMainUpdate{} rule.  The server can use this resource to update their own view
to the new state (\DbCopyUpdate{}).

The connection and disconnection predicates for the client and the server ensure
that the physical and logical databases are connected to the bridge resources.
In the case of a compromise, the client and the server can have inconsistent
views of the logical database, in which case we do not require the bridge
resources to be present.

Finally, to prove the specifications of \Cref{fig:key-value-spec}, we use RPC
invariants to inform the server about which operations are performed on the
logical state.  We leverage the fact that the RPC abstraction can be used to
transfer resources, which allows the client to send $\dbUpdate$ resources to
keep the server synchronized.  When the server receives these messages, it
synchronizes its copy of the logical state and applies the corresponding
operations to maintain its invariant.  For example, the invariants for storing
or loading a value are shown in \Cref{fig:key-value-assertions}.  In particular,
the response invariant for loading a value guarantees that the value $t_2$ in
the response is the correct value associated with the key $t_1$ sent in the
request.  Here, we make use of the fact that the invariant for the response of
the load request can mention the queried key $t_1$; otherwise, the client
wouldn't be able to tell that the value $t_2$ corresponds to $t_1$, since the
key does not appear in the response.

\section{Implementation}
\label{sec:implementation}

We implemented Cryptis as a Rocq library~\cite{Coq} with the Iris
framework~\cite{JungKJBBD18} and used it to verify the main examples of the
paper.  Iris allows defining expressive concurrent separation logics, with
support for higher-order ghost state, invariants and more.  Cryptis inherits
those features from Iris, and since they are orthogonal to the reasoning
patterns supported by Cryptis, it is possible to compose protocols with other
concurrent programs and reason about their behavior without compromising the
soundness of the logic.  Though the model of Iris is quite complex, most of this
complexity is shielded from the user; moreover, thanks to its generic
\emph{adequacy theorem}, it is possible to relate Iris proofs to the plain
operational semantics of the language.  Finally, Iris comes with an interactive
proof mode~\cite{KrebbersTB17}, which greatly simplifies the verification of
programs using the logic.

Rather than formalizing the Cryptis programming language from scratch, we
implemented it as a library in HeapLang, the default programming language used
in Iris developments.  We developed a small library of HeapLang programs to help
manipulating lists and other data structures. The resulting language differs in
a few respects compared to our paper presentation.  First, we formalized
cryptographic terms as a separate type from HeapLang values, and rely on an
explicit function to encode terms as values.  Thanks to this encoding, we can
ensure that Diffie-Hellman terms are normalized so that their intended notion of
equality coincides with equality in the Rocq logic, similar to some encodings of
quotient types in type theory~\cite{Cohen13}.  We implemented nonces as heap
locations, which allowed us to reuse much of the location infrastructure.  In
particular, our term metadata feature was inspired by a similar feature for
HeapLang locations, and its implementation uses metadata tokens associated with
locations to keep track of which terms are fresh and have not had any metadata
tokens generated for them.  Encoding nonces in terms of heap locations is
well-suited for reasoning about protocols in the symbolic model, but it is not
meant to be taken too literally---in particular, real cryptographic protocols
need to send messages over the wire as bit strings, and it is not reasonable to
expect that attackers that have access to the network at that level comply with
the representation constraints that we impose.

One difference between our implementation and our paper presentation is that we
have assumed that term variables always range over terms that have been
previously generated.  In Rocq, we cannot impose this restriction easily, so
instead we have a separate \verb!minted! predicate that ensures that every nonce
that appears in a term has been previously allocated.

Finally, when we proved results about programs and games, we assumed that the
attacker is implicitly running in the background. In our implementation,
instead, the attacker is explicitly initialized.  It allocates a list for
storing all the messages sent through the network, and launches a separate
thread that nondeterministically generates fresh terms or applies cryptographic
operations to other terms available to the attacker.  We maintain an invariant
that only public messages appear on this list.  When someone tries to receive a
message, the attacker nondeterministically chooses one of the messages that it
has seen or produced and returns that message to the user.

To give an idea of the effort involved in Cryptis, \Cref{fig:statistics} shows
the size of our development and case studies.  The ``Cryptis Core'' row
encompasses the Cryptis logic, the HeapLang libraries for manipulating terms and
their specifications.  The figures reported for case studies are broken down in
lines of code for the HeapLang implementation, lines of code for proofs of the
Cryptis specifications (aggregated with specifications and auxiliary
definitions), and lines of code for defining and proving the security of games.
We also include the time spent to compile the code with parallel compilation on
Rocq 9.0 running on an Ubuntu 24.04 laptop with an Intel i7-1185G7 3.00GHz with
eight cores and 15GiB of RAM.  These statistics show that the effort required by
Cryptis is comparable to other advanced tools for modular protocol verification,
such as DY*~\cite{Bhargavan21}.

\begin{figure}
  \centering
  \small
  \input{figures/table}
  \caption{Code statistics.}
  \Description{Code statistics}
  \label{fig:statistics}
\end{figure}

\section{Related Work}
\label{sec:related-work}

\paragraph*{Verification of message-passing or distributed applications}

Recent years have seen the introduction of several tools for reasoning about
distributed systems and message-passing concurrency, such as
Disel~\cite{SergeyWT18}, Actris~\cite{HinrichsenBK20,HinrichsenLKB21}, or
Aneris~\cite{Krogh-Jespersen20,GondelmanHPTB23}.  One common limitation of these
tools is that they assume a fairly reliable communication model.  For example,
Actris assumes that messages cannot be dropped, duplicated or tampered with,
whereas Aneris assumes that messages cannot be tampered with.  By contrast,
Cryptis allows us to reason about programs running over an adversarial network.
On the other hand, some of these tools have been designed to reason about more
challenging idioms of message-passing programming than what we currently handle
in Cryptis.  For example, Actris uses session types to reason about the
communication between agents.  In future work, we would like to bring together
these two lines of research, by developing an extension of Cryptis that
integrates the reasoning principles identified by these and other tools for
reasoning about distributed systems (e.g., integrate session types with our
communication abstraction).

\paragraph*{Tools for protocol verification}

There is a vast literature on techniques for verifying cryptographic protocols;
see \textcite{BarbosaBBBCLP19} for a recent survey.  One line of work in this
wider landscape focuses on verifying the absence of memory-safety violations or
other low-level bugs in protocol
implementations~\cite{ErbsenPGSC19,PolubelovaBPBFK20}.  Ruling out such bugs is
crucial to ensure the security of a protocol, but does not suffice to establish
all of its required integrity and confidentiality guarantees.
Such properties are usually analyzed with specialized tools.

Protocol-analysis tools strike a balance between many requirements, such as
expressiveness, convenience, and scalability.  In one corner of the design
space, we have automated solvers such as ProVerif~\cite{Blanchet01},
Tamarin~\cite{MeierSCB13} and CPSA~\cite{DoghmiGT07}, which favor convenience
over expressiveness and scalability, but which are nonetheless powerful enough
to analyze several real-world protocols.  Cryptis explores a different set of
trade-offs: limited support for push-button automation, but a more scalable
analysis and an expressive assertion language, which enables the reuse of
protocol specifications within proofs of larger systems.

Many other tools settle for similar trade-offs.  The work that is the closest to
ours is DY*~\cite{Bhargavan21}. DY* is a state-of-the-art F* library for
protocol verification that has been used to verify various protocols, such as
the Signal messaging protocol~\cite{Bhargavan21} or the ACME
protocol~\cite{BhargavanB0HKSW21}.  Like Cryptis, DY* is based on the symbolic
model of cryptography and emphasizes expressiveness, allowing users to state and
verify complex properties.  The verification is carried out manually, with
partial automation support---in the case of DY*, by leveraging the F* type
system and SMT solvers.  DY* was designed to enable the extraction of executable
code from protocol models---a feature that would be crucial for making protocol
implementations more reliable, but that Cryptis currently lacks.

Regarding our focus, the reuse of protocol proofs to verify larger systems,
there are several differences between the two tools.  First, DY* is not based on
separation logic, so it does not support the several verification idioms that
rely on it.  Second, DY* has a restricted state model: agents can keep
serialized long-term state associated with individual sessions, but do not have
access to common primitives for manipulating heap data structures.
Finally, DY* is based on a sequential semantics, whereas programs in Cryptis run
with a nondeterministic scheduler.  Since cryptographic protocols are concurrent
systems, they can only be modeled in DY* by manually following a rigid coding
discipline.  We must decompose each protocol into a series of protocol actions,
where each action is defined in a separate function whose behavior does not rely
on scheduling nondeterminism.  If several actions are inadvertently combined
into one function, its specification might hold only for a restricted choice of
interleavings, leaving the protocol open to attack.

These differences suggest that it would be difficult replicate in DY* the same
type of proof reuse that Cryptis supports.  Consider a system such as our
key-value store.  In DY*, it would be impossible to formulate succinct,
self-contained specifications for the client wrappers, such as those of
\Cref{fig:key-value-spec}.  On the one hand, we would need to decompose each
wrapper into several atomic actions to factor out local and network-wide
scheduling nondeterminism.  On the other hand, our specifications make crucial
use of connectives that have no analogue outside of separation logic (e.g., a
points-to connective to model the state of the store).

Another difference between the two tools, orthogonal to the goal of end-to-end
verification, lies in the support for compositionality. DY* enables
compositionality through a \emph{layered approach}~\cite{BhargavanBHKPSWW23}: a
protocol can be defined as a composition of several layers, where each layer
specifies disjointness conditions that should be respected by other components,
as well as predicates that need to be proved by its clients when using a
cryptographic primitive.  For example, if a component $C$ uses an encryption key
is shared with other components, we must specify all encrypted messages that $C$
is allowed to manipulate, and the other components cannot manipulate such
messages in ways that conflict with what $C$ expects.  The message invariants of
Cryptis play a similar role, but sacrifice some generality in return for ease of
use: protocols can be composed automatically if they rely on disjoint message
tags, a phenomenon that has been observed several times in the
literature~\cite{CiobacaC10, ArapinisCD12, ArapinisCD15, Maffei05,
  BugliesiFM04-ccs, BugliesiFM04, AndovaCGMMR08}.  Tag disjointness only needs
to be checked once, when declaring tag invariants; by contrast, disjointness
conditions in DY* need to be checked on every call to a cryptographic primitive.
This means that protocol composition in Cryptis can be obtained as a simple
consequence of the general composition rules of separation logic.

On a parallel line of work, several authors have proposed ways of integrating
symbolic cryptography within automated program analyzers for separation
logic~\cite{Vanspauwen015,ArquintSM023}.  These proposals aim to verify that
protocol implementations are free of memory safety violations while also
conforming to their expected confidentiality and integrity guarantees.  The work
of \textcite{ArquintSM023} was the first to demonstrate that separation-logic
resources are useful to reason about protocol security beyond just
memory-related bugs, by using special freshness resources to prove that
protocols satisfy \emph{injective agreement} (the absence of replay attacks).
Our metadata tokens enables similar, but more general, reasoning patterns.  In
particular, they can be used to prevent replay attacks at the application level,
by guaranteeing that reliable connections deliver each message only once.

Besides relying on a larger trusted computing base, one important difference
with respect to Cryptis is that these works do not attempt to reuse proofs of
protocol correctness to reason about larger systems.  It is not obvious how
these proposals could be leveraged to support this kind of reasoning. Our
reliable connection abstraction, for instance, uses the term metadata feature of
Cryptis to enable the transfer of resources through an authenticated connection,
a feature that plays a crucial role in the verification of our key-value store.

Looking beyond symbolic cryptography, several works have been developed to
reason about protocols in the computational
model~\cite{AbateHRMWHMS21,BartheDGKSS13,StoughtonCGQ22,GancherGSDP23}.  The
computational model is more realistic than the symbolic model on which Cryptis
is based, since it assumes that attackers have the power to manipulate messages
as raw bitstrings, without being confined to a limited API of operations.  On
the other hand, dealing with such attackers requires more detailed reasoning,
which means that such tools have difficulty scaling beyond individual
cryptographic primitives or simple protocols.

\paragraph*{Specification of authentication}

Most works on protocol verification view a protocol as a means for agents to
agree on their identities, protocol parameters, session keys, or the order of
events during the execution~\cite{Lowe97a, Bhargavan21, DattaM0S11, GordonJ03,
  ArquintSM023, Blanchet01, MeierSCB13}.  For example, if an initiator $I$
authenticates with a responder $R$, we might want to guarantee that $R$ was
indeed running at some point in the past, that it was running and accepted to
connect with $I$ specifically, or that it accepted to start a unique session
with $I$ that corresponds to the session key that they exchanged~\cite{Lowe97a}.

Cryptis demonstrates that agreeing on identities and on the contents of
exchanged messages is crucial when reusing a protocol.  For example, when a
key-value store receives a database operation, it must know which agent sent
this request to apply the operation to the correct database; when the client
receives the response, it must keep track of which key was queried to know which
value will be returned.  This aspect of authentication is implicit in the
specifications of key-exchange protocols in Cryptis, which allow us to determine
the identity of participants based on the exchanged session key.  On the other
hand, we have not found an instance where the the exact ordering of events in an
authentication handshake could be leveraged to reason about a larger system that
uses a protocol.  This allowed us to define the Cryptis logic without a global
trace of events, which is often used in other works in this space to state
authentication specifications~\cite{Bhargavan21, ArquintSM023}.

\section{Conclusion and Future Work}
\label{sec:conclusion}

We presented Cryptis, an Iris extension for symbolic cryptographic reasoning.
As we demonstrated throughout the paper, Cryptis allows us to reduce the
correctness of distributed systems verified in separation logic to elementary
assumptions embodied by the symbolic model of cryptography, without the need for
baking in a stronger (and less realistic) communication model.  The integration
of cryptographic reasoning allows us to evaluate how the correctness of a system
is affected by compromising cryptographic material such as a long-term private
key, going beyond what standard specifications in separation logic provide.
Thanks to the adequacy of the Iris logic, which Cryptis inherits, these
correctness results can be understood in rather concrete terms, via security
games that rely only on the operational semantics of the underlying programming
language.

Like related tools~\cite{Bhargavan21}, Cryptis is limited to single executions.
This can be restrictive for security, since many secrecy properties talk about
pairs of executions (e.g. indistinguishability).  We plan to lift this
restriction in the future, drawing inspiration from Sumii and Pierce's work on
reasoning about sealing via logical relations~\cite{SumiiP07/seal,SumiiP03}, as
well as recent work on a relational variant of Iris~\cite{FruminKB18}.  Another
avenue for strengthening the logic would be to incorporate probabilistic
properties and the computational model of cryptography.  Recent work shows that
probabilistic reasoning can benefit from separation logic~\cite{BartheHL20}, and
we believe that these developments could be naturally incorporated to our
setting.  Finally, we plan to extend the tool to encompass more protocols by
adding more cryptographic primitives (e.g. group inverses would allow us to
analyze the recent OPAQUE protocol~\cite{JareckiKX18}).

\aaa{Need to have a look at \cite{CanettiH11}, which seems to combine symbolic
  analysis with UC guarantees}

\printbibliography

\end{document}

%% file: figures/system-structure.tex
\begin{tikzpicture}
  [component/.style={
     draw,
     align=center,
     minimum size=2cm,
     font=\tiny
   },
   internal/.style={component,circle},
   top level/.style={component}]

   \node[internal] (authentication)
   {Authenticated\\key exchange\\(\Cref{sec:nsl,sec:iso})};

   \node[internal,right=of authentication] (connections)
   {Reliable\\connections\\(\Cref{sec:reliable-connections})};

   \node[internal,right=of connections] (rpc)
   {Remote\\procedure calls\\(\Cref{sec:rpc})};

   \node[internal,above=of authentication] (data structures)
   {Data\\structures};

   \node[top level] (server) at (data structures -| connections)
   {Server};

   \node[top level] (client) at (server -| rpc)
   {Client\\wrappers};

   \draw[->] (authentication) -- (connections);
   \draw[->] (connections) -- (rpc);
   \draw[->] (rpc) -- (client);
   \draw[->] (rpc) -- (server.south);
   \draw[->] (data structures) -- (server);
\end{tikzpicture}


%% file: figures/table.tex
\begin{tabular}{r|l|l|l|l|l}
Component & Impl. (loc) & Proofs (loc) & Game (loc) & Total (loc) & Wall-clock time (s) \\\hline
Cryptis Core & --- & --- & --- & 9679 & 237 \\ 
NSL (\Cref{sec:nsl}) & 54 & 224 & 254 & 532 & 40 \\ 
ISO (\Cref{sec:iso}) & 59 & 809 & 348 & 1241 & 56 \\ 
Connections (\Cref{sec:reliable-connections}) & 79 & 546 & --- & 638 & 28 \\
RPC (\Cref{sec:rpc}) & 52 & 499 & --- & 561 & 26 \\ 
Store (\Cref{sec:key-value}) & 154 & 1361 & 164 & 1687 & 78 \\
\end{tabular}